\documentstyle[aps,prl,epsfig]{revtex}
\begin{document}
%\draft
\title{Josephson vortices and the Meissner effect in stacked junctions and layered
superconductors: Exact analytical results}
\author{Sergey V. Kuplevakhsky}
\address{Department of Physics, Kharkov National University,\\
61077 Kharkov, Ukraine}
\date{\today}
\maketitle

\begin{abstract}
We present an exact mathematical description of Josephson vortices and of
the Meissner effect in periodic thin-layer superconductor/insulator
structures with an arbitrary number of identical junctions $N-1$ ($2\leq
N<\infty $, where $N$ is the number of superconducting layers) in terms of
localized solutions to a system of differential equations for phase
differences. We establish a general criterion of the existence of localized
solutions. We show that Meissner solutions are characterized by several
Josephson lengths $\lambda _{Ji}$ ($\frac N2$ lengths for even $N$, and $%
\frac{N-1}2$ lengths for odd $N$). We derive an exact expression for the
superheating field of the Meissner state, $H_s$, as an explicit function of $%
N$. For Josephson vortices, we find two basically different types of
topological solutions: ''vortex-plane'' solutions and incoherent vortex
solutions. Thermodynamically stable ''vortex-plane'' solutions represent a
chain of $N-1$ vortices (one vortex per each insulating layer). They are
characterized by the same set of $\lambda $$_{Ji}$ as the Meissner
solutions. We obtain exact analytical expressions for their self-energy and
for the lower critical field $H_{c1}$. Incoherent vortex solutions comprise
solutions with $k<N-1$ vortices and different vortex-antivortex
configurations. In contrast to the ''vortex-plane'' solutions, they prove to
be thermodynamically unstable, and their spatial dependence is
characterized, in general, by $N-1$ length scales. As an illustration, we
analyze 1-4-Josephson-junction stacks and investigate a transition to the
layered superconductor limit ($N\rightarrow \infty $).
\end{abstract}


\section{Introduction}

We present a rigorous mathematical examination of the problem of Josephson
vortices and of the Meissner effect in thin-layer Josephson-junction stacks
and layered superconductors, with a static external magnetic field ${\bf H}$
applied parallel to the layers (along the $z$ axis, see Fig. 1.) We consider
periodic systems composed of an arbitrary number $N-1$ of identical
superconductor/insulator (S/I) junctions ($2\leq N<\infty $, where $N$ is
the number of S-layers, with $x$ being the layering axis). Our starting
point is the microscopic Gibbs free-energy functional derived in Ref. \onlinecite
{K99}. Mathematical structure of this functional is analogous to that of
the phenomenological Lawrence-Doniach model. \cite{LD} Thus, the treatment of
our paper fully applies to the latter model as well.

Mathematically, both Josephson vortices in moderate fields $\left| {\bf H}%
\right| $ and the inhomogeneous Meissner state are described by solutions of
a system of nonlinear second-order differential equations for phase
differences, $\phi _n$, with square-integrable first-order derivatives.\cite
{K99,K00} (For brevity, we call here such solutions ''localized''). Our
approach is substantially based on the observation of a nontrivial property
of the differential equations for $\phi _n$: We show that the problem of
finding localized solutions for $\phi _n$ can be reduced to solving a
standard initial value problem. Using this key mathematical result, we
establish an exact criterion of the existence of localized solutions. The
existence criterion, in turn, allows us to obtain a complete classification
of physical localized solutions. We have found three types of such
solutions: Meissner solutions, topological ''vortex-plane'' solutions,\cite
{K99,K00} and topological incoherent vortex solutions.

Meissner solutions are localized near the side boundaries $y=-L$ and $y=L$.
In contrast to the well-known single-junction case,\cite{K70,BP82} the
Meissner solutions in stacks with $N\geq 4$ turn out to be characterized by
several Josephson lengths $\lambda _{Ji}$ ($\frac N2$ lengths for even $N$,
and $\frac{N-1}2$ lengths for odd $N$). The Meissner solutions persist up to
a certain superheating field of the Meissner phase, $H_s$. We derive an
exact expression for $H_s$ as an explicit function of $N$. We show that the
field $H_s$ simultaneously determines the penetration field for ''vortex
planes''. (See below.)

Thermodynamically stable ''vortex-plane'' solutions represent a chain of $%
N-1 $ Josephson vortices (one vortex per each I-layer), positioned in the
symmetry plane $y=0$. These solutions are uniquely determined by the
vortex-penetration conditions at $\left| {\bf H}\right| =H_s$. They are
characterized by the same set of $\lambda _{Ji}$ as the Meissner solutions.
Such solutions were previously obtained for infinite ($N=\infty $) layered
superconductors.\cite{K99,K00} Under the name of the ''coherent mode'' or
the ''in-phase mode'' they are well-known in double-junction stacks.\cite
{SBP93,SAUK96,GGU96,KW97} For $4\leq N<\infty $, the existence of the
''coherent mode'' was predicted in Ref. \onlinecite{SBP93}. (The authors of Ref.
\onlinecite{SBP93} specially emphasized the importance of this mode for practical
applications.) Besides giving a proof of the existence and stability of the
''vortex planes'' in the general case $2\leq N<\infty $, we derive exact
analytical expressions for their self-energy $E_v$ and for the lower
critical field $H_{c1}$.

Incoherent vortex solutions comprise single-vortex solutions, vortex
solutions with $2\leq k<N-1$ vortices in the plane $y=0,$ as well as
different vortex-antivortex configurations. All such solutions satisfy the
existence criterion. However, in contrast to the ''vortex-plane'' solutions,
they prove to be thermodynamically unstable and do not meet the
vortex-penetration conditions at any $\left| {\bf H}\right| \not =0$. It
should be noted that the single-vortex solutions obtained in this paper have
no resemblance to hypothetical Abrikosov-type vortices, introduced without
proper mathematical justification in some previous publications.\cite
{B73,CC90} Besides being thermodynamically unstable, the actual
single-vortex solutions are not uniquely determined by asymptotic boundary
conditions. They are accompanied by singular phase-difference distribution
in all $N-1$ junctions, and their spatial dependence is characterized, in
general, by $N-1$ length scales.

Section II of the paper is devoted to exact mathematical formulation of the
problem. In section III, we derive all major physical and mathematical
results sketched above. The general consideration of this section is
illustrated by several concrete examples in section IV. In particular, we
analyze 1-4-junction stacks and investigate a transition to the
layered-superconductor limit ($N\rightarrow \infty $). The obtained results
are discussed in section V. Appendices A-C contain some additional
mathematics, relevant to the subject of our study.

\section{Formulation of the problem}

We begin by writing down the microscopic Gibbs free-energy functional\cite
{K99} of a periodic structure consisting of alternating $N$ superconducting
(S) and $N-1$ insulating (I) layers ($2\leq N<\infty $):
\[
\Omega \left[ f_{n},\phi _{n},\frac{d\varphi _{n}}{dy},A_{x},A_{y};H\right] =%
\frac{H_{c}^{2}(T)}{4\pi }aW_{z}\left[ \sum_{n=0}^{N-1}\int\limits_{-L}^{L}dy%
\left[ -f_{n}^{2}(y)+\frac{1}{2}f_{n}^{4}(y)\right. \right.
\]
\[
\left. +\zeta ^{2}(T)\left( \frac{df_{n}(y)}{dy}\right) ^{2}+\zeta ^{2}(T)%
\left[ \frac{d\varphi _{n}(x,y)}{dy}-2eA_{y}(np,y)\right] ^{2}f_{n}^{2}(y)%
\right]
\]
\[
+\sum_{n=1}^{N-1}\int\limits_{-L}^{L}dy\left[ \frac{r(T)}{2}\left[
f_{n-1}^{2}(y)+f_{n}^{2}(y)-2f_{n}(y)f_{n-1}(y)\cos \left[ \phi
_{n}(y)-2e\int\limits_{(n-1)p}^{np}dxA_{x}(x,y)\right] \right] \right.
\]
\begin{equation}
\left. +\left. \frac{4e^{2}\zeta ^{2}(T)\lambda ^{2}(T)}{a}%
\int\limits_{(n-1)p}^{np}dx\left[ H(x,y)-H\right] ^{2}\right] \right] ,
\label{1.1}
\end{equation}
\[
r(T)\equiv \frac{\zeta ^{2}(T)\alpha }{a\xi _{0}},
\]
\[
\alpha \equiv \frac{3\pi ^{2}}{7\zeta (3)}\int\limits_{0}^{1}dttD(t)\ll 1,
\]
\[
\phi _{n}(y)=\varphi _{n}(y)-\varphi _{n-1}(y).
\]
Here $\hbar =c=1$; $a$ is the S-layer thickness; $p$ is the period, and $%
W_{z}$ is the length of the structure in the $z$ direction ($%
W_{z}\rightarrow \infty $); the length of the structure in the $y$ direction
is $W_{y}=2L$; $f_{n}(y)$ [$0\leq f_{n}(y)\leq 1$] and $\varphi _{n}(y)$
are, respectively, the reduced modulus and the phase of the pair potential $%
\Delta _{n}(y)$ in the $n$th superconducting layer:
\[
\Delta _{n}(y)=\Delta (T)f_{n}(y)\exp \varphi _{n}(y),
\]
with $\Delta (T)$ being the microscopic gap at temperature $T$; $\xi _{0}$
is the BCS coherence length; $\zeta (T)$ and $\lambda (T)$ are,
respectively, the Ginzburg-Landau (GL) coherence length and the penetration
depth; $D(\cos \theta )$ is the incidence-angle-dependent tunneling
probability of the I-layer between two successive S-layers; $H_{c}(T)$ is
the thermodynamic critical field; ${\bf A}=(A_{x},A_{y},0)$ is the vector
potential. The local magnetic field ${\bf H}(x,y)=\left[ 0,0,H(x,y)\right] $
obeys the Maxwell equation
\[
H(x,y)=\frac{\partial A_{y}(x,y)}{\partial x}-\frac{\partial A_{x}(x,y)}{%
\partial y}
\]
with boundary conditions
\[
H\left( 0,y\right) =H\left( \left( N-1\right) p,y\right) =H,\quad y\in \left[
-L,L\right] ,
\]
\[
H\left( x,\pm L\right) =H,\quad x\in \left[ 0,\left( N-1\right) p\right] ,
\]
where $H$ is a static external magnetic field applied along the $z$ axis.
(See Fig. 1.) The sum of the first three phase- and field-independent terms
on the right-hand side of (\ref{1.1}) represents the condensation energy.
The fourth term is the kinetic energy of the intralayer currents. The last
two terms are the Josephson energy and the field energy, respectively.

Expression (\ref{1.1}) is valid under the conditions
\begin{equation}  \label{1.9}
\frac{T_{c0}-T}{T_{c0}}\ll 1,
\end{equation}
\begin{equation}  \label{1.10}
\xi _0\ll a,
\end{equation}
\begin{equation}  \label{1.11}
a\ll \min \left\{ \zeta (T),\lambda (T),\alpha ^{-1}\xi _0\right\} ,
\end{equation}
\begin{equation}  \label{1.12}
a\ll p.
\end{equation}
Conditions (\ref{1.9}) ($T_{c0}$ is the critical temperature of an isolated
S-layer) and (\ref{1.10}) ensure the applicability of the GL-type expansion
within each S-layer. Condition (\ref{1.11}) corresponds to the thin S-layer
limit, whereas condition (\ref{1.12}) is employed here for the sake of
mathematical simplicity only. Being a first-order expansion in $a/p$,
equation (\ref{1.1}) applies in fields $\left| H\right| \ll H_{c2}$, where $%
H_{c2}$ is the upper critical field.

Mathematical treatment of the functionals of the type (\ref{1.1}) is
described in full detail in Ref. \onlinecite{K00}, section III: One minimizes (\ref
{1.1}) with respect to $f_n$ and $A_x$, $A_y$, imposes the gauge $A_x=0$ and
eliminates $A_y$ by integration. The result is a closed, complete set of
coupled nonlinear mean-field equations for the reduced modulus of the pair
potential $f_n$ and the phase differences $\phi _n$, together with relations
for all physical quantities of interest. To simplify mathematical analysis
of the mean-field equations, we introduce dimensionless units by
\[
\frac xp\rightarrow x,
\]
\[
\frac y{\lambda _{J\infty }}\rightarrow y,
\]
\[
\frac H{H_{s\infty }}\rightarrow H,
\]
where the quantities on the left-hand side are dimensional, with $\lambda
_{J\infty }=\left( 8\pi ej_0p\right) ^{-1/2}$ being the Josephson
penetration depth ($j_0$ is the density of the Josephson current in a single
junction with thick electrodes) and $H_{s\infty }=\left( ep\lambda _{J\infty
}\right) ^{-1}$ being the superheating (penetration) field of the infinite
layered superconductor.\cite{K99,K00} In our dimensionless units, for
example, the flux quantum is $\Phi _0=\pi $, and the lower critical field of
the infinite layered superconductor\cite{K99,K00} is $H_{c1\infty }=\frac 2%
\pi $.

In the dimensionless form, the mean-field equations for $f_n$ and $\phi _n$
read
\[
f_0(y)-f_0^3(y)=r(T)\left[ \frac{\epsilon ^2}2\frac{d^2f_0(y)}{dy^2}+\frac{2%
\left[ H-H_1(y)\right] ^2}{\epsilon ^2f_0^3(y)}+\frac 12\left[
f_0(y)-f_1(y)\cos \phi _1(y)\right] \right] ,
\]
\[
f_n(y)-f_n^3(y)=r(T)\left[ \frac{\epsilon ^2}2\frac{d^2f_n(y)}{dy^2}+\frac{2%
\left[ H_n(y)-H_{n+1}(y)\right] ^2}{\epsilon ^2f_n^3(y)}\right.
\]
\begin{equation}  \label{1.16}
\left. +\frac 12\left[ 2f_n(y)-f_{n+1}(y)\cos \phi _{n+1}(y)-f_{n-1}(y)\cos
\phi _n(y)\right] \right] ,\quad 1\leq n\leq N-2,
\end{equation}
\[
f_{N-1}(y)-f_{N-1}^3(y)=r(T)\left[ \frac{\epsilon ^2}2\frac{d^2f_{N-1}(y)}{%
dy^2}+\frac{2\left[ H-H_{N-1}(y)\right] ^2}{\epsilon ^2f_{N-1}^3(y)}\right.
\]
\[
\left. +\frac 12\left[ f_{N-1}(y)-f_{N-2}(y)\cos \phi _{N-1}(y)\right] %
\right] ;
\]
\begin{equation}  \label{1.17}
\frac{df_n}{dy}\left( \pm L\right) =0,\quad 0\leq n\leq N-1;
\end{equation}
\[
\frac 1{f_n^2(y)}\left[ H_{n+1}(y)-H_n(y)\right] -\frac 1{f_{n-1}^2(y)}\left[
H_n(y)-H_{n-1}(y)\right] -\epsilon ^2H_n(y)
\]
\begin{equation}  \label{1.18}
=-\frac{\epsilon ^2}2\frac{d\phi _n(y)}{dy},\quad 1\leq n\leq N-1,
\end{equation}
\begin{equation}  \label{1.19}
H_0(y)=H_N(y)=H,
\end{equation}
\begin{equation}  \label{1.20}
\frac{d\phi _n}{dy}\left( \pm L\right) =2H,\quad 1\leq n\leq N-1,
\end{equation}
where
\begin{equation}  \label{1.21}
\epsilon \equiv \frac{\sqrt{ap}}\lambda <1,
\end{equation}
and the local magnetic field in the $n$th insulating layer ($n-1<x<n$) is
given by
\[
H_n(y)=\frac 12\int\limits_{-L}^yduf_n(u)f_{n-1}(u)\sin \phi _n(u)+H
\]
\begin{equation}  \label{1.22}
=\frac 12\int\limits_L^yduf_n(u)f_{n-1}(u)\sin \phi _n(u)+H.
\end{equation}
Note that although in most physical situations $\epsilon \ll 1$, for the
sake of mathematical generality, in this paper, $\epsilon $ is supposed to
satisfy only the weak inequality (\ref{1.21}). Because of the obvious
property $f_n(y)=f_n(-y)$, equation (\ref{1.22}) implies that the phase
differences $\phi _n$ meet the condition
\begin{equation}  \label{1.23}
\phi _n(y)=-\phi _n(-y)+0\text{mod}2\pi .
\end{equation}

The dimensionless Gibbs free energy $\Omega (H)$, normalized via the relation
\[
\frac{4\pi \Omega (H)}{H_c^2(T)a\lambda _{J\infty }W_z}\rightarrow \Omega
(H),
\]
in terms of the mean-field quantities $f_n$, $\phi _n$ and $H_n(y)$ has the
form
\[
\Omega (H)=\sum_{n=0}^{N-1}\int\limits_{-L}^Ldy\left[ -f_n^2(y)+\frac 12%
f_n^4(y)+\frac{r(T)\epsilon ^2}2\left( \frac{df_n(y)}{dy}\right) ^2\right.
\]
\[
\left. +\frac{2r(T)}{\epsilon ^2f_n^2(y)}\left[ H_{n+1}(y)-H_n(y)\right]
^2+2r(T)\left[ H_n(y)-H\right] ^2\right]
\]
\begin{equation}  \label{1.24}
+\frac{r(T)}2\sum_{n=1}^{N-1}\int\limits_{-L}^Ldy\left[
f_{n-1}^2(y)+f_n^2(y)-2f_n(y)f_{n-1}(y)\cos \phi _n(y)\right] .
\end{equation}
Here, the two terms in the second line on the right-hand side are the
kinetic energy of the intralayer currents and the field energy,
respectively. The intralayer current in the $n$th S-layer $J_n(y)$
(normalized to $H_{s\infty }$) and the density of the Josephson current
between the $n$th and the $(n-1)$th S-layers $j_{n,n-1}(y)$ (normalized to $%
j_0$) are given by
\begin{equation}  \label{1.25}
J_n(y)=\frac 1{4\pi }\left[ H_n(y)-H_{n+1}(y)\right] ,\quad 0\leq n\leq N-1,
\end{equation}
and
\begin{equation}  \label{1.26}
j_{n,n-1}(y)=2\frac{dH_n(y)}{dy}=f_n(y)f_{n-1}(y)\sin \phi _n(y),\quad 1\leq
n\leq N-1,
\end{equation}
respectively. Relations (\ref{1.16})-(\ref{1.26}) provide a complete,
self-consistent description of the thin-layer periodic S/I structure in the
temperature range (\ref{1.9}) and in fields $\left| H\right| \ll H_{c2}$.

In this paper, we will be interested in physical solutions for $\phi _n$
with a square-integrable first-order derivative, localized within a spatial
range of order unity. (For brevity, we call such solutions ''localized''.)
Therefore we assume the condition
\begin{equation}  \label{1.27}
L\gg 1.
\end{equation}
Moreover, we assume that the temperature range satisfies the condition of
the weak-coupling limit
\begin{equation}  \label{1.28}
r(T)\ll 1.
\end{equation}
One can obtain a perturbative solution for $f_n$ and $\phi _n$ up to any
desired order in $r(T)$, starting from the zero-order solution to (\ref{1.16}%
), (\ref{1.17}),
\begin{equation}  \label{1.29}
f_n=1,
\end{equation}
and the zero-order equations for $\phi _n$,
\begin{equation}  \label{1.30}
H_{n+1}(y)-\left( 2+\epsilon ^2\right) H_n(y)+H_{n-1}(y)=-\frac{\epsilon ^2}2%
\frac{d\phi _n(y)}{dy},\quad 1\leq n\leq N-1,
\end{equation}
where $H_n(y)$ are given by (\ref{1.22}) with $f_n=1$ and satisfy the
boundary conditions (\ref{1.19}). For most applications, it is sufficient to
consider expressions for physical quantities only in leading order in $r(T)$%
. Thus, for example, substituting (\ref{1.29}) and the solution of (\ref
{1.30}) into (\ref{1.24}) immediately yields a first-order expansion for the
Gibbs free energy, because first-order corrections to the
condensation-energy term cancel out.

A detailed mathematical analysis of Eqs. (\ref{1.30}) is the subject of
section III. Here we point out that these equations can be transformed into
a very useful for application form by solving for $H_n(y)$ (see Appendix A
for mathematical details):
\begin{equation}  \label{1.31}
H_n(y)=h_n(y)+H_n,
\end{equation}
\begin{equation}  \label{1.32}
h_n(y)=\frac{\epsilon ^2}2\sum_{m=1}^{N-1}G(n,m)\frac{d\phi _m(y) }{dy},
\end{equation}
\begin{equation}  \label{1.33}
H_n=\frac{H\left( \mu ^{-n}+\mu ^{-N+n}-\mu ^n-\mu ^{N-n}\right) }{\mu
^{-N}-\mu ^N},
\end{equation}
where $G(n,m)$ are given by (\ref{b.9}), and $\mu $ is given by (\ref{b.5}).
By (\ref{b.8}), and (\ref{1.20}), (\ref{b.13}), expression (\ref{1.31})
explicitly satisfies boundary conditions (\ref{1.19}) and $H_n(\pm L)=H$.
Moreover, the $y$-independent quantities $H_n$ in (\ref{1.31}) have clear
physical meaning: Being solutions of (\ref{1.30}) with $\frac{d\phi _m(y)}{dy%
}\equiv 0$, they describe distribution of the local magnetic field within
I-layers in the homogeneous Meissner state (see section III of Ref. \onlinecite
{K00}). Also note that $H_n=H_{N-n}$, which is a reflection of the symmetry
of the problem. [By comparison, in an infinite layered superconductor $%
H_n\equiv 0$, and $\sum\limits_{m=1}^{N-1}G(n,m)\ldots \rightarrow
\sum\limits_{m=-\infty }^{+\infty }G_\infty (n,m)\ldots ,$ where $G_\infty
(n,m)$ are defined by (\ref{b.15}).]

In addition, we point out that equations of the popular phenomenological
Lawrence-Doniach model\cite{LD} also can be reduced to the dimensionless
form (\ref{1.16})-(\ref{1.20}), (\ref{1.22})-(\ref{1.26}), with $r(T)$ being
a phenomenological parameter, $\epsilon \equiv \frac p\lambda $, and $\Omega
(H)$ normalized to $\frac{H_c^2(T)p\lambda _{J\infty }W_z}{4\pi }$. (See
Ref. \onlinecite{K00} for more details.) Thus, all the consideration of this paper
fully applies to the Lawrence-Doniach model as well.

\section{Major results}

\subsection{The criterion of the existence of localized solutions}

By differentiation with respect to $y$, integrodifferential equations (\ref
{1.30}) reduce to a system of $N-1$ ordinary nonlinear second-order
differential equations
\[
\frac{d^2\phi _1(y)}{dy^2}=\frac 1{\epsilon ^2}\left[ \left( 2+\epsilon
^2\right) \sin \phi _1(y)-\sin \phi _2(y)\right] ,
\]
\[
\frac{d^2\phi _n(y)}{dy^2}=\frac 1{\epsilon ^2}\left[ \left( 2+\epsilon
^2\right) \sin \phi _n(y)-\sin \phi _{n+1}(y)-\sin \phi _{n-1}(y)\right]
,\quad 2\leq n\leq N-2,
\]
\begin{equation}  \label{1.34}
\frac{d^2\phi _{N-1}(y)}{dy^2}=\frac 1{\epsilon ^2}\left[ \left( 2+\epsilon
^2\right) \sin \phi _{N-1}(y)-\sin \phi _{N-2}(y)\right]
\end{equation}
with boundary conditions (\ref{1.20}).

Consider Eqs. (\ref{1.34}) on the whole axis $-\infty <y<+\infty $. Two
simple properties of (\ref{1.34}) are quite obvious: If $\phi _n(y)$ ($1\leq
n\leq N-1$) is a solution, the functions $\bar \phi _n(y)$ given by
\begin{equation}  \label{1.35}
\bar \phi _n(y)=\phi _n(y)+2\pi k\quad \text{(}k\text{ is an integer),}
\end{equation}
and
\begin{equation}  \label{1.36}
\bar \phi _n(y)=\phi _n(y+c)\quad \text{(}c\text{ is an arbitrary constant)}
\end{equation}
are also solutions. [The latter is a result of the fact that $y$ does not
enter explicitly the right-hand side of (\ref{1.34}).] Our conclusions about
the existence of localized solutions to (\ref{1.34}) will be substantially
based on another key property, which we formulate as a lemma:

{\bf Lemma}. Consider an arbitrary interval $I=\left[ L_1,L_2\right] $ and $%
y_0\in I$. The initial value problem for Eqs. (\ref{1.34}) with arbitrary
initial conditions $\phi _n(y_0)=\alpha _n$, $\frac{d\phi _n}{dy}(y_0)=\beta
_n$ has a unique solution in the {\it whole} interval $I$. This solution has
continuous derivatives with respect to $y$ of arbitrary order and
continuously depends on the initial data. (For the proof of the {\bf Lemma},
see Appendix B.)

It is worth noting that the existence and uniqueness of a smooth solution to
the initial value problem in the {\it whole} interval $I$ is rather
nontrivial for {\it nonlinear} differential equations: For such equations,
theorems of existence and uniqueness are usually valid only locally, in the
neighborhood of initial data.\cite{T61} In our case, global character of the
solution and its infinite differentiability are ensured by the fact that $%
\phi _n$ enter the right-hand side of Eqs. (\ref{1.34}) only as arguments of
the sine. Note that because of the arbitrariness of the interval $I$, the
solution can be uniquely continued onto the whole axis $-\infty <$$y<+\infty
$.\cite{T61} Now we will show that the problem of finding localized
solutions to (\ref{1.34}) can be reduced to the standard initial value
problem.

Differentiating (\ref{1.31}) with respect to $y$ yields
\begin{equation}  \label{1.37}
\sin \phi _n(y)=\epsilon ^2\sum_{m=1}^{N-1}G(n,m)\frac{d^2\phi _m(y)}{dy^2}.
\end{equation}
Multiplying (\ref{1.37}) by $\frac{d\phi _n(y)}{dy}$, summing over the layer
index $n$ with the use of (\ref{b.11}) and performing integration, we arrive
at the first integral of Eqs. (\ref{1.34}):
\begin{equation}  \label{1.38}
C-\sum_{n=1}^{N-1}\cos \phi _n(y)=\frac{\epsilon ^2}2\sum_{n=1}^{N-1}%
\sum_{m=1}^{N-1}G(n,m)\frac{d\phi _n(y)}{dy}\frac{d\phi _m(y)}{dy},
\end{equation}
where $C$ is the constant of integration. Now let us choose an arbitrary
point $y_0\in \left[ -L,L\right] $, where $L$ is sufficiently large [see the
condition (\ref{1.27})]. We are looking for localized solutions of Eqs. (\ref
{1.34}) that in the region
\begin{equation}  \label{1.39}
\lambda _{\max }\ll \left| y-y_0\right| ,
\end{equation}
where $\lambda _{\max }$ is determined by the maximum positive eigenvalue of
the symmetric matrix ${\bf \tilde G}(n,m)$ (see Appendix A), satisfy the
asymptotic conditions
\begin{equation}  \label{1.40}
\phi _n(y)=0\text{mod}2\pi +o(1),
\end{equation}
\begin{equation}  \label{1.41}
\frac{d\phi _n(y)}{dy}=o(1)
\end{equation}
for any $1\leq n\leq N-1$. By inserting (\ref{1.40}) and (\ref{1.41}) into (%
\ref{1.38}), we establish the value of the constant of integration: $C=N-1$.
Thus Eq. (\ref{1.38}) becomes
\begin{equation}  \label{1.42}
\sum_{n=1}^{N-1}\sin {}^2\frac{\phi _n(y)}2=\frac{\epsilon ^2}4%
\sum_{n=1}^{N-1}\sum_{m=1}^{N-1}G(n,m)\frac{d\phi _n(y)}{dy}\frac{d\phi _m(y)%
}{dy}.
\end{equation}
Substituting initial values $\phi _n(y_0)=\alpha _n$, $\frac{d\phi _n}{dy}%
(y_0)=\beta _n$ into (\ref{1.42}), we obtain the general {\it criterion of
the existence }of localized solutions to (\ref{1.34}):
\begin{equation}  \label{1.43}
\sum_{n=1}^{N-1}\sin {}^2\frac{\alpha _n}2=\frac{\epsilon ^2}4%
\sum_{n=1}^{N-1}\sum_{m=1}^{N-1}G(n,m)\beta _n\beta _m.
\end{equation}
Indeed, the {\bf Lemma} guarantees the existence, uniqueness and
differentiability of a solution for arbitrary $\alpha _n$ and $\beta _n$ in
the whole interval $\left[ -L,L\right] $. Owing to our special choice of the
constant of integration in (\ref{1.42}), the solution determined by $\alpha
_n$ and $\beta _n$ obeying (\ref{1.43}) will necessarily satisfy asymptotic
conditions (\ref{1.40}), (\ref{1.41}) in the region (\ref{1.39}). Moreover,
this solution will automatically satisfy the conditions
\begin{equation}  \label{1.44}
\frac{d^k\phi _n(y)}{dy^k}=o(1)
\end{equation}
in the region (\ref{1.39}) for any $1\leq n\leq N-1$ and $2\leq k$, which
can be verified by repeated differentiation of (\ref{1.37}) and application
of (\ref{1.40}), (\ref{1.41}). Below, we apply the criterion (\ref{1.43}) to
concrete physical situations.

\subsection{Meissner solutions. The superheating (penetration) field $H_s$}

Let $H>0$ for definiteness. Consider an intermediate length scale $R$ such
that
\begin{equation}  \label{1.45}
1\ll R<\frac L2.
\end{equation}
(See Fig. 1.) The Meissner boundary value problem is specified by the
boundary conditions (\ref{1.20}) and
\begin{equation}  \label{1.46}
\phi _n\left( \pm (L-R)\right) =o(1),\quad \frac{d\phi _n}{dy}\left( \pm
(L-R)\right) =o(1)
\end{equation}
for any $1\leq n\leq N-1$. For $-L\leq y<-L+R$, we write using (\ref{1.22}):
\begin{equation}  \label{1.47}
H_n(y)=\frac 12\int\limits_{-L+R}^ydu\sin \phi _n(u)+\frac 12%
\int\limits_{-L}^{-L+R}du\sin \phi _n(u)+H.
\end{equation}
According to (\ref{1.31}), we make the identification
\begin{equation}  \label{1.48}
h_n(y)=\frac 12\int\limits_{-L+R}^ydu\sin \phi _n(u),
\end{equation}
\begin{equation}  \label{1.49}
H_n=\frac 12\int\limits_{-L}^{-L+R}dy\sin \phi _n(y)+H,
\end{equation}
where $h_n(y)$ stands for the field penetrating through the $y=-L$
interface. Using the fact that $H_n<H$, and $H_n=H_{N-n}$, we establish the
following important properties:
\begin{equation}  \label{1.50}
-\pi \leq \phi _n(y)<0,
\end{equation}
\begin{equation}  \label{1.51}
\phi _n(y)=\phi _{N-n}(y),\quad h_n(y)=h_{N-n}(y),\quad H_n(y)=H_{N-n}(y).
\end{equation}
Relations (\ref{1.51}) are a result of the symmetry of the problem. They
imply that the number of independent equations describing the Meissner
solution is $\frac N2$ for even $N$ and $\frac{N-1}2$ for odd $N$. The
solution for $\phi _n(y)$ in the region $L-R<y\leq L$ can be obtained from
the solution in the region $-L\leq y<-L+R$ using the property
\begin{equation}  \label{1.51.1}
\phi _n(y)=-\phi _n(-y),
\end{equation}
resulting from the general relation (\ref{1.23}). In the region $-R<y<R$, $%
\phi _n\equiv 0$, and we have
\begin{equation}  \label{1.52}
H_n(y)=H_n.
\end{equation}

Now we apply relation (\ref{1.43}) for $y_0\equiv -L$. In view of the
boundary conditions $\frac{d\phi _n}{dy}(-L)\equiv \beta _n=2H$ [see (\ref
{1.20})], we get
\begin{equation}  \label{1.53}
\frac 1{N-1}\sum_{n=1}^{N-1}\sin {}^2\frac{\alpha _n}2=\frac{H^2 }{H_s^2},
\end{equation}
where
\begin{equation}  \label{1.54}
H_s=\left[ 1-\frac{\left( 2\sqrt{1+\frac{\epsilon ^2}4}-\epsilon \right)
\left( 1-\mu ^{N-1}\right) }{\epsilon \left( N-1\right) \left( 1+\mu
^{N-1}\right) }\right] ^{-\frac 12}.
\end{equation}

Physical interpretation of the quantity $H_s$ is straightforward. The
maximum value of the left-hand side of (\ref{1.53}) is unity, which
corresponds to $H=H_s$ on the right-hand side. Hence $H_s$ is the maximum
external field for which a Meissner solution is still possible, i.e. the
{\it superheating} field of the Meissner state.\cite{K70,BP82} Moreover, the
maximum of the left-hand side is achieved when all $\alpha _n\equiv \phi
_n(-L)=-\pi $, which, by (\ref{1.51.1}), gives $\phi _n(L)-\phi _n(-L)=2\pi $
for {\it all} $n$. The total flux of the penetrating field at $H=H_s,$%
\[
\Phi
=\sum_{n=1}^{N-1}\int\limits_{-L}^{-L+R}dyh_n(y)+\sum_{n=1}^{N-1}\int%
\limits_{L-R}^Ldyh_n(y)
\]
\[
=\pi \left( N-1\right) \left[ 1-\frac{2\sqrt{1+\frac{\epsilon ^2}4}-\epsilon
}{\epsilon \left( N-1\right) }\frac{1-\mu ^{N-1}}{1+\mu ^N}\right] ,
\]
is exactly equal to the total flux carried by a ''vortex plane'', i.e. a
chain of $N-1$ vortices positioned in the plane $y=0$. [See Eq. (\ref{1.60})
below.] These conditions correspond\cite{BP82,K99} to {\it simultaneous} and
{\it coherent} penetration of Josephson vortices into all the junctions.
Therefore $H_s$ can be also regarded as the {\it penetration} field for a
''vortex plane''. Note that $H_s$ for $N<\infty $ is always higher than the
penetration field of an infinite layered superconductor $H_{s\infty }=1$.
\cite{K99,K00}

Thus, the Meissner boundary value problem, Eqs. (\ref{1.20}) and (\ref{1.46}%
), for $H=H_s$ reduces to the initial value problem with $\alpha _n\equiv
\phi _n(-L)=-\pi $ and $\beta _n\equiv \frac{d\phi _n(-L)}{dy}=2H_s$ ($1\leq
n\leq N-1$). Let $\phi _n^{H_s}(y)$ be the corresponding unique solution. By
continuous dependence of a solution of the initial value problem on initial
data (see the {\bf Lemma}), we can argue that the existence of the unique
Meissner solution $\phi _n^{H_s}(y)$ at $H=H_s$ guarantees the existence of
a unique Meissner solution at any $H<H_s$ with a certain set of $\alpha _n$ (%
$-\pi <\alpha _n<0$) satisfying (\ref{1.53}). In other words, we have proved
that the Meissner boundary value problem, Eqs. (\ref{1.20}) and (\ref{1.46}%
), has a unique solution for any $H\leq H_s$.

Finally, we want to emphasize one more important property of Meissner
solutions in ($N-1$)-junction periodic S/I structures that was not realized
in previous publications.\cite{BF92} Owing to the fact that these solutions
satisfy a system of coupled second-order differential equations ($\frac N2$
equations for even $N,$ and $\frac{N-1}2$ equations for odd $N$), they are
necessarily characterized by {\it several} Josephson lengths $\lambda _{Ji}$
($\frac N2$ lengths for even $N,$ and $\frac{N-1}2$ lengths for odd $N$), in
contrast to a single junction with only one $\lambda _J$. All $\lambda _{Ji}$
are determined by positive eigenvalues of the symmetric matrix ${\bf \tilde G%
}(n,m)$. (See Appendix A.)

\subsection{Vortex-plane solutions. The lower critical field $H_{c1}$}

By a vortex-plane solution we understand a chain of $N-1$ Josephson vortices
(one vortex per each I-layer) positioned at $y=0$. Analogously, an
antivortex-plane solution is a chain of $N-1$ Josephson antivortices (one
antivortex per each I-layer). Such solutions are characterized by the
symmetry
\begin{equation}  \label{1.55}
\phi _n(y)=\pm 2\pi -\phi _n(-y)
\end{equation}
[see (\ref{1.23})] and asymptotic boundary conditions\cite{R82}
\begin{equation}  \label{1.56}
\phi _n(-R)=o(1),\quad \phi _n(R)=\pm 2\pi +o(1),
\end{equation}
\begin{equation}  \label{1.57}
\frac{d^k\phi _n(\pm R)}{dy^k}=o(1)
\end{equation}
for {\it all} $1\leq n\leq N-1$ and {\it any} $k\geq 1.$ [The ''plus'' sign
in (\ref{1.55}) and (\ref{1.56}) corresponds to a vortex plane in fields $%
H>0 $, whereas the ''minus'' sign corresponds to an antivortex plane in
fields $H<0$.] The existence and uniqueness of these solutions follows
immediately from the {\bf Lemma}, the criterion (\ref{1.43}) and the results
of the previous subsection. Indeed, by (\ref{1.55}), a vortex (antivortex)
plane satisfies the conditions $\alpha _n\equiv $$\phi _n(0)=\pm \pi $ for
all $n$. Under these conditions, the left-hand side of (\ref{1.43}) reaches
its maximum. As shown in the previous subsection, the maximum condition
corresponds to the unique choice $\beta _n\equiv \frac{d\phi _n}{dy}(0)=\pm
2H_s$ on the right-hand side of (\ref{1.43}), consistent with the
vortex-penetration conditions (see the next paragraph). Hence the initial
values
\begin{equation}  \label{1.57.1}
\alpha _n\equiv \phi _n(0)=\pm \pi ,\quad \beta _n\equiv \frac{d\phi _n}{dy}%
(0)=\pm 2H_s,\quad 1\leq n\leq N-1,
\end{equation}
determine a unique localized solution in the region $-R<y<R$ that
automatically meets the asymptotic boundary conditions (\ref{1.56}), (\ref
{1.57}).

Thermodynamic stability of vortex-plane (antivortex-plane) solutions is
ensured by the fact that they satisfy the vortex-penetration conditions for $%
H=\pm H_s$. Let $H>0$ for definiteness. (In what follows, we consider only
vortex planes. The discussion of antivortex planes is quite analogous.) A
vortex-plane solution can be constructed from the Meissner solution $\phi
_n^{H_s}$ in the region $-L\leq y<-L+R$, discussed in the previous
subsection. Indeed, using the properties (\ref{1.35}), (\ref{1.36}), we
obtain a solution
\[
\bar \phi _n(y)=\phi _n^{H_s}(y-L)+2\pi
\]
in the region $0\leq y<R$ that satisfies the initial conditions $\bar \phi
_n(0)=\pi $, $\frac{d\bar \phi _n}{dy}(0)=2H_s$. This solution can be
continued\cite{T61} into the region $-R<y\leq 0$. By the uniqueness of a
solution to the initial value problem, the obtained solution coincides with
the vortex-plane solution in the interval $-R<y<R$.

The local magnetic field in the presence of a vortex plane is given by
general relations (\ref{1.31}) and (\ref{1.47})-(\ref{1.49}), where $h_n(y)$
[with $y\in (-R,R)$] should be interpreted as the magnetic field induced by
the vortex plane itself. Owing to the vortex-plane initial conditions $\beta
_n\equiv $$\frac{d\phi _n}{dy}(0)=2H_s$,
\begin{equation}  \label{1.61}
h_n(0)=H_s\left[ 1-\frac{\mu ^{-n}+\mu ^{-N+n}-\mu ^n-\mu ^{N-n} }{\mu
^{-N}-\mu ^N}\right] .
\end{equation}
(See Fig. 2.) The symmetry relations (\ref{1.51}) apply to vortex-plane
solutions too. Thus, the number of independent equations describing a vortex
plane is $\frac N2$ for even $N,$ and $\frac{N-1}2$ for odd $N$, as in the
case of Meissner solutions. Consequently, spatial dependence of vortex-plane
solutions is characterized by the same set of $\lambda _{Ji}$ as that of the
Meissner solutions.

The flux $\Phi _n$ through the $n$th I-layer can be found using (\ref{1.32}%
), (\ref{1.56}) and (\ref{b.13}):
\begin{equation}  \label{1.59}
\Phi _n=\int\limits_{-R}^Rdyh_n(y)=\pi \left[ 1-\frac{\mu ^{-n}+\mu
^{-N+n}-\mu ^n-\mu ^{N-n}}{\mu ^{-N}-\mu ^N}\right] .
\end{equation}
The total flux carried by a vortex plane is
\begin{equation}  \label{1.60}
\Phi =\sum_{n=1}^{N-1}\Phi _n=\pi \left( N-1\right) \left[ 1- \frac{2\sqrt{1+%
\frac{\epsilon ^2}4}-\epsilon }{\epsilon \left( N-1\right) } \frac{1-\mu
^{N-1}}{1+\mu ^N}\right] .
\end{equation}
Note that in contrast to Josephson junctions with thick electrodes\cite{BP82}
and infinite layered superconductors,\cite{K99,K00} the flux carried by a
Josephson vortex in a finite thin-layer S/I structure {\it is not quantized}
and is always smaller than the flux quantum $\Phi _0=\pi $. (This fact has
been already pointed out in Ref. \onlinecite{KW97}.)

To determine the thermodynamic {\it lower critical }field $H_{c1}$ at which
the vortex-plane solutions become energetically favorable, we must calculate
the difference between the Gibbs free energy in the presence of a single
vortex plane, $\Omega _v(H)$, and the Gibbs free energy of the homogeneous
Meissner state, $\Omega _M(H)$ [the sum of phase-independent terms in (\ref
{1.24})]. Substituting (\ref{1.31})-(\ref{1.33}) into (\ref{1.24}) and using
(\ref{b.7}), (\ref{1.42}), in first order in $r(T)\ll 1$, we obtain:
\[
\Omega _v(H)-\Omega _M(H)
\]
\begin{equation}  \label{1.62}
=r(T)\left[ \epsilon
^2\sum_{n=1}^{N-1}\sum_{m=1}^{N-1}G(n,m)\int\limits_{-R}^Rdy\frac{d\phi
_n(y) }{dy}\frac{d\phi _m(y)}{dy}-4\Phi H\right] ,
\end{equation}
where the total flux $\Phi $ is given by (\ref{1.60}). The first term on the
right-hand side of (\ref{1.62}) should be interpreted as the self-energy of
the vortex plane:
\[
E_v=r(T)\epsilon
^2\sum_{n=1}^{N-1}\sum_{m=1}^{N-1}G(n,m)\int\limits_{-R}^Rdy \frac{d\phi
_n(y)}{dy}\frac{d\phi _m(y)}{dy}
\]
\[
=4r(T)\int\limits_{-R}^Rdy\sum_{n=1}^{N-1}\sin {}^2\frac{\phi _n(y)}2
\]
\begin{equation}  \label{1.63}
=r(T)\sum_{n=1}^{N-1}\int\limits_{-R}^Rdy\left[ \frac{\epsilon ^2}2%
\sum_{m=1}^{N-1}G(n,m)\frac{d\phi _n(y)}{dy}\frac{d\phi _m(y)}{dy}+1-\cos
\phi _n(y)\right] .
\end{equation}
Note that $E_v$ is exactly twice the energy of the Josephson currents in (%
\ref{1.24}). Besides, formulas (\ref{1.62}) and (\ref{1.63}), with
corresponding reinterpretation of $\Omega _v(H)$, $\Phi $ and $E_v$, also
hold for incoherent vortex solutions considered in the next subsection. They
also apply to an infinite layered superconductor, taking account of the
substitution $\sum\limits_{m,n=1}^{N-1}G(n,m)\ldots \rightarrow
\sum\limits_{m,n=-\infty }^{+\infty }G_\infty (n,m)\ldots $, where $G_\infty
(n,m)$ are defined by (\ref{b.15}). In this latter case, the self-energy
should be additionally minimized with respect to the {\it phases} $\varphi
_n $, which immediately yields the exact solution\cite{K99,K00} with $\phi
_n(y)=\phi _{n+1}(y)\equiv \phi (y)$ and $\lambda _{J\infty }=1$. [In the
case $N<\infty $, the minimization with respect to $\varphi _n$ is not
allowed by the boundary conditions (\ref{1.19}).\cite{K00}]

From (\ref{1.62}), we get:
\begin{equation}  \label{1.64}
H_{c1}=\frac{E_v}{4r(T)\Phi }.
\end{equation}
For thermodynamically stable solutions, we must necessarily have $H_{c1}<H_s$%
. It is straightforward to verify that this condition is met by the
vortex-plane solutions. Using (\ref{1.37}), (\ref{1.55}) (with the ''plus''
sign), the initial values $\beta _n=2H_s$ and integrating by parts, we
convert $E_v$ into the form
\begin{equation}  \label{1.65}
E_v=2r(T)\left[ \sum_{n=1}^{N-1}\int_R^0dy\phi _n(y)\sin \phi _n(y)-2H_s\Phi %
\right] .
\end{equation}
The first term on the right-hand side of (\ref{1.65}) is positive, because
in the region $0\leq y<R$ all $\phi _n$ satisfy the relation $\pi \leq \phi
_n<2\pi $. By the use of (\ref{1.37}) and (\ref{b.14}), we obtain the
following strict inequalities:
\[
2H_s\Phi <\sum_{n=1}^{N-1}\int_R^0dy\phi _n(y)\sin \phi _n(y)<4H_s\Phi ,
\]
\[
0<E_v<4r(T)H_s\Phi .
\]
Hence,
\[
0<H_{c1}<H_s,
\]
as anticipated. Note that in all special cases admitting exact analytical
solutions ($N=\infty $,\cite{K99,K00} and $N=2,3$, see section IV), $H_{c1}=%
\frac 2\pi H_s$. Finally, we want to point out that in contrast to
vortex-plane solutions in infinite layered superconductors,\cite{K99,K00}
where $H_n(y)=H_{n+1}(y)=H(y)$ and $J_n=0$ for all $n$, in finite structures
the intralayer currents $J_n$, in general, are not equal to zero, as can be
easily seen from (\ref{1.25}). Only for even number of junctions ($N$ is
odd), in the central S-layer $J_{\frac{N-1}2}=0$, by the symmetry (\ref{1.51}%
).

\subsection{Single-vortex solutions and other localized incoherent vortex
solutions}

A single Josephson vortex positioned in the $l$th I-layer at $y=0$ obeys
symmetry relations
\begin{equation}  \label{1.68}
\phi _l(y)=2\pi -\phi _l(-y);\quad \phi _n(y)=-\phi _n(-y),\quad n\neq l,
\end{equation}
[see (\ref{1.23})] and asymptotic boundary conditions\cite{R82}
\begin{equation}  \label{1.69}
\phi _l(-R)=o(1),\quad \phi _l(R)=2\pi +o(1),
\end{equation}
\begin{equation}  \label{1.70}
\phi _n(\pm R)=o(1),\quad n\neq l,
\end{equation}
\begin{equation}  \label{1.71}
\frac{d^k\phi _n(\pm R)}{dy^k}=o(1),\quad \text{for all 1}\leq n\leq N-1%
\text{, and any }k\geq 1.
\end{equation}
Moreover, $\frac{dh_n(y)}{dy}>0$ in the region $-R<y<0$, and $\frac{dh_n(y)}{%
dy}<0$ in the region $0<y<R$. Hence, $\phi _n$ must satisfy the relations
\begin{equation}  \label{1.72}
0<\phi _n(y)<\pi ,\text{ }y\in \left( -R,0\right) ;\quad -\pi <\phi _n(y)<0,%
\text{ }y\in \left( 0,R\right) ,\quad \text{for }n\neq l,
\end{equation}
[see (\ref{1.48}) for $y\in (-R,R)$] and the initial conditions
\begin{equation}  \label{1.73}
\alpha _l\equiv \phi _l(0)=\pi ;\quad \alpha _n\equiv \phi _n(0)=0,\quad
n\neq l,
\end{equation}
\begin{equation}  \label{1.74}
\beta _l\equiv \frac{d\phi _l(0)}{dy}>0;\quad \beta _n\equiv \frac{d\phi
_n(0)}{dy}<0,\quad n\neq l.
\end{equation}
A necessary condition of the existence of such solutions is provided by the
general criterion (\ref{1.43}) and has the form
\begin{equation}  \label{1.75}
\frac{\epsilon ^2}4\sum_{n=1}^{N-1}\sum_{m=1}^{N-1}G(n,m)\beta _n\beta _m=1.
\end{equation}
Relation (\ref{1.75}) imposes only one constraint on $N-1$ quantities $\beta
_n$ and can be satisfied by different sets of $\beta _n$. Therefore, in
contrast to the vortex-plane problem, a solution to the single-vortex
problem for any given vortex position $l$ is {\it not unique}. The
determination of the optimum set of $\beta _n$ may require an additional
examination of the vortex self-energy (see section III.B).

Besides failing to meet the requirement of uniqueness, single-vortex
solutions, in general, break the symmetry (\ref{1.51}), inherent to the
original integrodifferential equations (\ref{1.18}), (\ref{1.30}) minimizing
the Gibbs free-energy functional (\ref{1.1}). (The only exclusion is a stack
with an odd number of junctions $N-1$ and $l=\frac N2$.) Even more important
is the fact that single-vortex configurations do not satisfy the
vortex-penetration conditions for any $H>0$. Indeed, starting from a
single-vortex solution in the region $-R<y<R$, we can construct a solution
in the region $-L\leq y<-L+R$ satisfying the Meissner boundary conditions (%
\ref{1.46}). (Compare with the reverse procedure of the construction of a
vortex-plane solution from the Meissner solution $\phi _n^{H_s},$ described
in the previous subsection.) However, the solution thus obtained will not
represent any physics, because it cannot meet the physical boundary
conditions $\frac{d\phi _n(-L)}{dy}\equiv \beta _n=2H$ for any $H>0$:
According to (\ref{1.74}), $\beta _l$ and $\beta _n$ with $n\not =l$ have
{\it different} signs. Physically, this means that isolated vortices cannot
penetrate the periodic S/I structure at any static $H>0$, and the
penetration field for {\it isolated vortices} cannot be defined. On the
basis of these observations, we conclude that, in contrast to vortex-plane
solutions, single-vortex solutions are{\it \ thermodynamically unstable} and
do not represent any solutions to (\ref{1.18}), (\ref{1.30}) for $H>0$. This
situation has a simple mathematical explanation. In contrast to differential
equations (\ref{1.34}), integrodifferential equations (\ref{1.18}), (\ref
{1.30}) explicitly contain the external magnetic field $H$. [See the
explicit expressions for $H_n(y)$, Eqs. (\ref{1.22}).] In the case of Eqs. (%
\ref{1.34}), the external field $H$ enters only via the boundary conditions (%
\ref{1.20}) at $y=\pm L$, whereas single-vortex configurations are required
to satisfy asymptotic boundary conditions (\ref{1.69})-(\ref{1.71}) at $%
y=\pm R$. It is therefore not surprising that Eqs. (\ref{1.34}), restricted
to the interval $(-R,R)\subset [-L,L]$, may possess redundant solutions that
do not satisfy Eqs. (\ref{1.18}), (\ref{1.30}). [Note that the exact
equations (\ref{1.18}), valid for arbitrary $r(T)$, do not, in general,
reduce to any differential equations, if $f_n(y)\not =$const.]

The flux through the $n$th I-layer due to the vortex in the $l$th I-layer
can be found using (\ref{1.32}) and (\ref{1.69}), (\ref{1.70}):
\begin{equation}  \label{1.76}
\Phi _n=\int\limits_{-R}^Rdyh_n(y)=\frac{\pi \epsilon }{2\sqrt{1+ \frac{%
\epsilon ^2}4}}\left[ \mu ^{\left| n-l\right| }-\frac{\mu ^n\left( \mu
^{l-N}-\mu ^{N-l}\right) +\mu ^{N-n}\left( \mu ^{-l}-\mu ^l\right) }{\mu
^{-N}-\mu ^N}\right] .
\end{equation}
The total flux carried by the vortex in the $l$th layer is
\begin{equation}  \label{1.77}
\Phi =\sum_{n=1}^{N-1}\Phi _n=\pi \left[ 1-\frac{\mu ^{-l}+\mu ^{-N+l}-\mu
^l-\mu ^{N-l}}{\mu ^{-N}-\mu ^N}\right] .
\end{equation}
For $N<\infty $, the total flux $\Phi $ is not quantized and is less than
the flux quantum $\Phi _0=\pi $. (See the previous subsection.) Note that
the total flux carried by a single vortex in the $l$th I-layer is exactly
equal to the flux through the $l$th I-layer in the case of a vortex plane,
given by Eq. (\ref{1.59}) with $n=l$.

The consideration of other localized incoherent vortex solutions to (\ref
{1.34}) (i.e., solutions with $2\leq k<N-1$ vortices in the plane $y=0$, and
vortex-antivortex configurations) can be done along the same lines. All
these solutions are thermodynamically unstable too. We want to underline
that our general conclusions about the form of static single-vortex
configurations completely agree with the results of numerical calculations
of Ref. \onlinecite{SBP93}. (See, in particular, Fig. 5 therein.) On the other
hand, these single-vortex configurations have no resemblance to hypothetical
Abrikosov-type vortices introduced without appropriate mathematical
justification in Refs. \onlinecite{B73,CC90}. Let alone thermodynamic instability,
the actual single-vortex solutions to (\ref{1.34}) are accompanied by
singular phase-difference distribution in {\it all} $N-1$ junctions,
satisfying (\ref{1.72})-(\ref{1.74}) and the existence condition (\ref{1.75}%
). Their spatial dependence is characterized, in general, by $N-1$ length
scales, which is inherent to localized solutions of a system of coupled
second-order differential equations. These intrinsic features of topological
solutions in discrete periodic S/I structures cannot be reproduced by any
imitation of Abrikosov vortices, typical of continuum type-II
superconductors. It is also worth reminding that Abrikosov vortices in the
London approximation are described by {\it linear} partial differential
equations,\cite{dG} whereas the ordinary differential equations (\ref{1.34})
are essentially {\it nonlinear}. The principle of superposition of solutions
is not valid for nonlinear equations. Unfortunately, this basic point is
sometimes disregarded in literature.\cite{BF92}

\section{Particular examples}

\subsection{A single thin-layer junction ($N=2$)}

In this simplest case, the only nonzero element of the matrix ${\bf G}(n,m)$%
, given by (\ref{b.9}), is
\begin{equation}  \label{4.1}
G(1,1)=\frac 1{2+\epsilon ^2}.
\end{equation}
By (\ref{1.37}), a single phase difference $\phi _1(y)\equiv \phi (y)$
satisfies the usual static sine-Gordon equation
\begin{equation}  \label{4.2}
\frac{d^2\phi (y)}{dy^2}=\frac 1{\lambda _J^2}\sin \phi (y),
\end{equation}
with the Josephson length\cite{K70}
\begin{equation}  \label{4.3}
\lambda _J=\frac \epsilon {\sqrt{2+\epsilon ^2}}.
\end{equation}
Note that $\lambda _J$, given by (\ref{4.3}), for $\epsilon \ll 1$ is much
smaller than the Josephson length of a single junction with thick
electrodes, which in our dimensionless units is $\lambda _{J0}=\sqrt{\frac p{%
2\lambda }}$.\cite{K70} From (\ref{1.33}) and (\ref{1.54}), we get the local
field in the homogeneous Meissner state
\begin{equation}  \label{4.4}
H_1=\frac{2H}{2+\epsilon ^2},
\end{equation}
and the superheating (penetration) field
\begin{equation}  \label{4.5}
H_s=\lambda _J^{-1}=\frac{\sqrt{2+\epsilon ^2}}\epsilon ,
\end{equation}
respectively. For $\epsilon \ll 1$, the superheating (penetration) field $%
H_s $, given by (\ref{4.5}), is much higher than the corresponding field\cite
{K70,BP82} of a single junction with thick electrodes $H_{s0}=\lambda _{J0}$.

\subsubsection{The Meissner solution}

For the fields $0\leq H\leq H_s=\frac{\sqrt{2+\epsilon ^2}}\epsilon $, the
Meissner solution in the region $-L\leq y<-L+R$ up to first order in $%
r(T)\ll 1$ is given by
\begin{equation}  \label{4.6}
\phi (y)=-4\arctan \frac{H\exp \left[ -\frac{\left( L+y\right) }{\lambda _J}%
\right] }{H_s+\sqrt{H_s^2-H^2}},
\end{equation}
\begin{equation}  \label{4.7}
H(y)\equiv H_1(y)=h(y)+H_1,
\end{equation}
\begin{equation}  \label{4.8}
h(y)\equiv h_1(y)=\frac{2\lambda _JH\left[ H_s+\sqrt{H_s^2-H^2}\right] \exp %
\left[ -\frac{\left( L+y\right) }{\lambda _J}\right] }{\left[ H_s+\sqrt{%
H_s^2-H^2}\right] ^2+H^2\exp \left[ -\frac{2\left( L+y\right) }{\lambda _J}%
\right] },
\end{equation}
\[
j(y)\equiv j_{1,0}(y)=-4H\left[ H_s+\sqrt{H_s^2-H^2}\right]
\]
\begin{equation}  \label{4.9}
\times \frac{\left[ \left[ H_s+\sqrt{H_s^2-H^2}\right] ^2-H^2\exp \left[ -%
\frac{2\left( L+y\right) }{\lambda _J}\right] \right] \exp \left[ - \frac{%
\left( L+y\right) }{\lambda _J}\right] }{\left[ \left[ H_s+\sqrt{H_s^2-H^2}%
\right] ^2+H^2\exp \left[ -\frac{2\left( L+y\right) }{\lambda _J}\right] %
\right] ^2},
\end{equation}
\begin{equation}  \label{4.10}
J(y)\equiv J_0(y)=J_1(y)=\frac 1{4\pi }\left[ H-H_1-h(y)\right] ,
\end{equation}
\begin{equation}  \label{4.11}
f(y)\equiv f_0(y)=f_1(y)=1-\frac{r(T)}2\left[ \lambda _J^{-2}h(y)+\frac 2{%
\epsilon ^2}\left[ h(y)+H_1-H\right] ^2\right] .
\end{equation}
The Meissner solution in the region $L-R<y\leq L$ can be obtained from (\ref
{4.6})-(\ref{4.11}) by means of the substitution $y\rightarrow -y$, $\phi
(y)\rightarrow -\phi (-y)$. In the region $-R<y<R$, the solution is
\begin{equation}  \label{4.12}
\phi (y)\equiv 0,\quad h(y)\equiv 0,\quad j(y)\equiv 0,
\end{equation}
\begin{equation}  \label{4.13}
H(y)=H_1,
\end{equation}
\begin{equation}  \label{4.14}
J(y)=\frac 1{4\pi }\left[ H-H_1\right] ,
\end{equation}
\begin{equation}  \label{4.15}
f(y)=1-\frac{r(T)}{\epsilon ^2}\left[ H-H_1\right] ^2.
\end{equation}

\subsubsection{The vortex solution}

In the region $-R<y<R,$ the vortex (antivortex) solution satisfies the
initial conditions $\alpha \equiv \phi (0)=\pi $, $\beta \equiv \frac{d\phi
(0)}{dy}=\pm 2H_s$ [Eq. (\ref{1.57.1})] and has the form
\begin{equation}  \label{4.16}
\phi (y)=\pm 4\arctan \exp \left[ \frac y{\lambda _J}\right] ,
\end{equation}
\begin{equation}  \label{4.17}
h(y)=\pm \lambda _J\cosh {}^{-1}\left[ \frac y{\lambda _J}\right] ,
\end{equation}
\begin{equation}  \label{4.18}
j(y)=\mp 2\cosh {}^{-2}\left[ \frac y{\lambda _J}\right] \sinh {}\left[
\frac y{\lambda _J}\right] .
\end{equation}
The quantities $H(y)$, $J(y)$ and $f(y)$ are given by (\ref{4.7}), (\ref
{4.10}) and (\ref{4.11}), respectively, with $h(y)$ taken from (\ref{4.17}).
Note that the field induced by a vortex at $y=0$, in agreement with (\ref
{1.61}), is
\[
h(0)=\lambda _J=\frac{\epsilon ^2H_s}{2+\epsilon ^2},
\]
and not $h(0)=H_s$, as in the case of a single junction with thick
electrodes.\cite{BP82} By inserting (\ref{4.1}) and (\ref{4.16}) into (\ref
{1.63}), we obtain the vortex self-energy:
\begin{equation}  \label{4.18.1}
E_v=8r(T)\frac \epsilon {\sqrt{2+\epsilon ^2}}.
\end{equation}
The vortex flux, according to (\ref{1.60}), is
\[
\Phi =\pi \frac{\epsilon ^2}{2+\epsilon ^2},
\]
and the lower critical field, by (\ref{1.64}), is
\begin{equation}  \label{4.19}
H_{c1}=\frac 2\pi H_s=\frac 2\pi \frac{\sqrt{2+\epsilon ^2}}\epsilon .
\end{equation}
Thus, for $\epsilon \ll 1$, the vortex flux $\Phi \ll \Phi _0=\pi $, and the
lower critical field (\ref{4.19}) is much larger than the corresponding
field of a single junction with thick electrodes $H_{c10}=\frac 2\pi \sqrt{%
\frac p{2\lambda }}$, in agreement with Ref. \onlinecite{KW97}.

\subsection{A double-junction stack ($N=3$)}

In the double-junction case, the nonzero matrix elements (\ref{b.9}) of $%
{\bf G}(n,m)$ are
\begin{equation}  \label{4.20}
G(1,1)=G(2,2)=\frac{2+\epsilon ^2}{\left( 2+\epsilon ^2\right) ^2-1},\quad
G(1,2)=G(2,1)=\frac 1{\left( 2+\epsilon ^2\right) ^2-1}.
\end{equation}
The corresponding $2\times 2$ matrix ${\bf \tilde G}(n,m)$ (see Appendix A)
has two positive eigenvalues: $\frac{\lambda _{J1}^2}{\epsilon ^2}$ and $%
\frac{\lambda _2^2}{\epsilon ^2}$, with the lengths
\begin{equation}  \label{4.21}
\lambda _{J1}=\frac \epsilon {\sqrt{1+\epsilon ^2}},\quad \lambda _2=\frac %
\epsilon {\sqrt{3+\epsilon ^2}}.
\end{equation}
According to (\ref{1.54}), the superheating (penetration ) field is
\begin{equation}  \label{4.22}
H_s=\lambda _{J1}^{-1}=\frac{\sqrt{1+\epsilon ^2}}\epsilon ,
\end{equation}
which is smaller than the corresponding single-junction value (\ref{4.5}),
in agreement with Ref. \onlinecite{GGU96}. The application of (\ref{1.33}) yields
the value of the local field in the homogeneous Meissner state:
\begin{equation}  \label{4.23}
H_1=H_2=\frac H{1+\epsilon ^2}.
\end{equation}

\subsubsection{The Meissner solution}

The Meissner solution in the fields $0\leq H\leq H_s=\frac{\sqrt{1+\epsilon
^2}}\epsilon $ obeys the symmetry relations (\ref{1.51}). The substitution
of
\begin{equation}  \label{4.23.1}
\phi _1(y)=\phi _2(y)\equiv \phi (y)
\end{equation}
into (\ref{1.37}), using (\ref{4.20}), yields
\begin{equation}  \label{4.24}
\frac{d^2\phi (y)}{dy^2}=\frac 1{\lambda _{J1}^2}\sin \phi (y).
\end{equation}
Thus, all the results of the single-junction case, Eqs. (\ref{4.6})-(\ref
{4.15}), apply if we substitute $\lambda _J\rightarrow \lambda _{J1}$ (note
that $\lambda _J<\lambda _{J1}$), make the identification
\[
\phi (y)\equiv \phi _1(y)=\phi _2(y),\quad H(y)\equiv H_1(y)=H_2(y),\quad
h(y)\equiv h_1(y)=h_2(y),
\]
\[
j(y)\equiv j_{1,0}(y)=j_{2,1}(y),\quad J(y)\equiv J_0(y)=J_2(y),\quad
f(y)\equiv f_0(y)=f_2(y),
\]
and take the values of $H_s$ and $H_1$ from (\ref{4.22}) and (\ref{4.23}),
respectively. Moreover,
\begin{equation}  \label{4.26}
J_1(y)=0,
\end{equation}
and
\begin{equation}  \label{4.27}
f_1(y)=1-\frac{r(T)}{\lambda _{J1}^2}h(y).
\end{equation}

\subsubsection{The vortex-plane solution}

In the region $-R<y<R,$ the vortex-plane (antivortex-plane) solution
describes two vortices (antivortices) [one vortex (antivortex) per I-layer]
and satisfies the initial conditions $\alpha _1\equiv \phi _1(0)=\pi $, $%
\alpha _2\equiv \phi _2(0)=\pi $, $\beta _1\equiv \frac{d\phi _1(0)}{dy}=\pm
2H_s$, $\beta _2\equiv \frac{d\phi _2(0)}{dy}=\pm 2H_s$ [Eq. (\ref{1.57.1}%
)]. By (\ref{1.51}), it obeys the symmetry (\ref{4.23.1}) and Eq. (\ref{4.24}%
), with
\[
\phi (y)=\pm 4\arctan \exp \left[ \frac y{\lambda _{J1}}\right] .
\]
Thus, explicit expressions for $h_1(y)=h_2(y)\equiv h(y)$ and $%
j_{1,0}(y)=j_{2,1}(y)\equiv j(y)$ can be obtained from single-junction Eqs. (%
\ref{4.17}), (\ref{4.18}), taking account of the substitution $\lambda
_J\rightarrow \lambda _{J1}$. The quantities $H_1(y)=H_2(y)\equiv H(y)$, $%
J_0(y)=J_2(y)\equiv J(y)$ and $f_0(y)=f_2(y)\equiv f(y)$ are given by (\ref
{4.7}), (\ref{4.10}) and (\ref{4.11}), respectively, with $H_1$ taken from (%
\ref{4.23}). For $J_1(y)$ and $f_1(y)$, we have (\ref{4.26}) and (\ref{4.27}%
), respectively.

The vortex-plane self-energy is
\begin{equation}  \label{4.28}
E_v=16r(T)\lambda _{J1}=16r(T)\frac \epsilon {\sqrt{1+\epsilon ^2}},
\end{equation}
and the flux is
\begin{equation}  \label{4.29}
\Phi =2\pi \frac{\epsilon ^2}{1+\epsilon ^2},
\end{equation}
which immediately leads to the lower critical field:
\[
H_{c1}=\frac 2\pi H_s=\frac 2\pi \frac{\sqrt{1+\epsilon ^2}}\epsilon .
\]
As can be seen by comparing (\ref{4.28}) with the single-junction expression
(\ref{4.18.1}), the energy per vortex in the double-junction stack is
higher. Finally, the field induced by the vortex plane at $y=0$, according
to (\ref{1.61}), is
\[
h(0)\equiv h_1(0)=h_2(0)=\frac{\epsilon ^2H_s}{1+\epsilon ^2}.
\]

\subsubsection{The vortex-antivortex solution}

As can be easily seen, equations (\ref{1.34}) with $N=3$ admit in the region
$-R<y<R$ another exact topological solution, namely an incoherent
vortex-antivortex solution
\begin{equation}  \label{4.31}
\phi _1(y)=-\phi _2(y)\equiv \phi (y),
\end{equation}
where $\phi (y)$ is given by
\[
\phi (y)=4\arctan \exp \left[ \frac y{\lambda _2}\right] .
\]
With $\alpha _1=\alpha _2=\pi $ and $\beta _1=-\beta _2=\pm 2\lambda _2^{-1}$%
, the vortex-antivortex solution explicitly satisfies the existence
criterion (\ref{1.43}). By (\ref{1.63}), the self-energy of the
vortex-antivortex solution is
\begin{equation}  \label{4.33}
E_{va}=16r(T)\lambda _2=16r(T)\frac \epsilon {\sqrt{3+\epsilon ^2}},
\end{equation}
which is lower than the self-energy of the vortex-plane solution (\ref{4.28}%
).

However, the vortex-antivortex solution does not satisfy integrodifferential
equations (\ref{1.30}) with $N=3$: These equations do not possess the
symmetry (\ref{4.31}) for any $H\not =0$. The vortex-penetration conditions
also cannot be met, because $\beta _1\beta _2<0$. Moreover, the Gibbs free
energy (\ref{1.24}) of the vortex-antivortex pair is always positive with
respect to the Gibbs free energy of the Meissner state: The flux $\Phi $
carried by this pair is exactly equal to zero. Thus, the static
vortex-antivortex solution is thermodynamically unstable, in agreement with
the general consideration in section III. As shown in Ref. \onlinecite{SBP93}, the
vortex-antivortex solution can be realized in the dynamic regime, in the
presence of an external current applied to the central S-layer.

\subsubsection{Single-vortex solutions}

Consider a configuration with a single vortex in one I-layer (say, with $n=2$%
) and no vortices in the other. The solution representing this configuration
does not possess any symmetry. As shown in Appendix C, the self-energy of a
single vortex, $E_{sv}$, satisfies the exact inequality $E_{va}<E_{sv}<E_v$,
where $E_v$ and $E_{va}$ are given by (\ref{4.28}) and (\ref{4.33}),
respectively. On the other hand, the total flux carried by a single vortex,
according to (\ref{1.77}) with $N=3$ and $l=2$, is
\[
\Phi =\pi \frac{\epsilon ^2}{1+\epsilon ^2},
\]
which is exactly half the total flux of the vortex plane (\ref{4.29}).
Hence, the Gibbs free energy of a single vortex is positive with respect to
that of the Meissner state for $H\leq H_{c1}=\frac 2\pi \frac{\sqrt{%
1+\epsilon ^2}}\epsilon $. This example clearly illustrates thermodynamic
instability of single-vortex solutions discussed in section III.

The two second-order differential equations describing the single-vortex
configuration can be reduced\cite{T61} to one fourth-order equation
\[
\frac{d^4\phi _2}{dy^4}+\frac{2+\epsilon ^2}{\epsilon ^2}\sin \phi _2\left(
\frac{d\phi _2}{dy}\right) ^2-\frac{2+\epsilon ^2}{\epsilon ^2}\cos \phi _2
\frac{d^2\phi _2}{dy^2}
\]
\[
-\frac{\left[ \frac{2+\epsilon ^2}{\epsilon ^2}\sin \phi _2-\frac{d^2\phi _2
}{dy^2}\right] \left[ \left( 2+\epsilon ^2\right) \cos \phi _2\frac{d\phi _2
}{dy}-\epsilon ^2\frac{d^3\phi _2}{dy^3}\right] ^2}{1-\left[ \left(
2+\epsilon ^2\right) \sin \phi _2-\epsilon ^2\frac{d^2\phi _2}{dy^2}\right]
^2}
\]
\begin{equation}  \label{4.34}
+\sqrt{1-\left[ \left( 2+\epsilon ^2\right) \sin \phi _2-\epsilon ^2\frac{%
d^2\phi _2}{dy^2}\right] ^2}\left[ \frac{\left( 2+\epsilon ^2\right) ^2-1}{%
\epsilon ^4}\sin \phi _2-\frac{2+\epsilon ^2}{\epsilon ^2}\frac{d^2\phi _2}{%
dy^2}\right] =0.
\end{equation}
The phase difference $\phi _1(y)$ can be found without any additional
integration from the relations
\[
\phi _1=\arcsin \left[ \left( 2+\epsilon ^2\right) \sin \phi _2-\epsilon ^2
\frac{d^2\phi _2}{dy^2}\right] ,\quad -\frac \pi 2\leq \phi _1\leq \frac \pi %
2,
\]
\[
\phi _1=-\arcsin \left[ \left( 2+\epsilon ^2\right) \sin \phi _2-\epsilon ^2
\frac{d^2\phi _2}{dy^2}\right] +\pi ,\quad \frac \pi 2<\phi _1<\pi ,
\]
\begin{equation}  \label{4.35}
\phi _1=-\arcsin \left[ \left( 2+\epsilon ^2\right) \sin \phi _2-\epsilon ^2%
\frac{d^2\phi _2}{dy^2}\right] -\pi ,\quad -\pi <\phi _1<-\frac \pi 2.
\end{equation}
The initial conditions for the vortex solution are given by the relations $%
\alpha _1\equiv \phi _1(0)=0$, $\alpha _2\equiv \phi _2(0)=\pi $, $\beta
_1\equiv \frac{d\phi _1(0)}{dy}<0$, $\beta _2\equiv \frac{d\phi _2(0)}{dy}>0$
and must satisfy the existence criterion (\ref{1.75}):
\begin{equation}  \label{4.36}
\left( 2+\epsilon ^2\right) \left( \beta _1^2+\beta _2^2\right) -2\left|
\beta _1\right| \beta _2=\frac{4\left[ \left( 2+\epsilon ^2\right) ^2-1%
\right] }{\epsilon ^2}.
\end{equation}
In view of the condition $\beta _2>0$, the appropriate solution of (\ref
{4.36}) for $\beta _2$ is
\[
\beta _2\equiv \frac{d\phi _2(0)}{dy}=\frac 1{2+\epsilon ^2}\left[ \left|
\beta _1\right| +\frac 1\epsilon \sqrt{\left[ \left( 2+\epsilon ^2\right)
^2-1\right] \left[ 4\left( 2+\epsilon ^2\right) -\epsilon ^2\beta _1^2\right]
}\right] ,
\]
\[
0<\left| \beta _1\right| \leq \frac{2\sqrt{2+\epsilon ^2}}\epsilon .
\]
By the symmetry (\ref{1.68}), $\frac{d^2\phi _2(0)}{dy^2}=0$. The initial
condition on $\frac{d^3\phi _2}{dy^3}$ can be obtained from the relation
\[
\frac{d^3\phi _2(y)}{dy^3}=\frac 1{\epsilon ^2}\left[ \left( 2+\epsilon
^2\right) \cos \phi _2(y)\frac{d\phi _2(y)}{dy}-\cos \phi _1(y)\frac{d\phi
_1(y)}{dy}\right]
\]
that follows directly from (\ref{1.34}):
\[
\frac{d^3\phi _2(0)}{dy^3}=-\frac 1{\epsilon ^3}\sqrt{\left[ \left(
2+\epsilon ^2\right) ^2-1\right] \left[ 4\left( 2+\epsilon ^2\right)
-\epsilon ^2\beta _1^2\right] }.
\]
Thus, we have obtained a complete formulation of the single-vortex problem
in terms of the standard initial value problem. In agreement with general
consideration of section III, the problem admits a family of single-vortex
solutions parameterized by $\left| \beta _1\right| $.

Unfortunately, there are no general methods of analytical integration of
nonlinear differential equations of order higher than two. However,
numerical integration of (\ref{4.34}) with the above-derived initial
conditions should pose no problem. Moreover, it is not difficult to obtain
asymptotics of the single-vortex solution in the region $\left| y\right| \gg
\lambda _{J1}$. For $y\ll -\lambda _{J1}$, equations (\ref{4.34}), (\ref
{4.35}) can be linearized:
\begin{equation}  \label{4.39}
\frac{d^4\phi _2}{dy^4}-\frac{2\left( 2+\epsilon ^2\right) }{\epsilon ^2}%
\frac{d^2\phi _2}{dy^2}+\frac{\left( 2+\epsilon ^2\right) ^2-1}{\epsilon ^4}%
\phi _2=0,
\end{equation}
\begin{equation}  \label{4.40}
\phi _1=\left( 2+\epsilon ^2\right) \phi _2-\epsilon ^2\frac{d^2\phi _2}{dy^2%
}.
\end{equation}
The solution of (\ref{4.39}), (\ref{4.40}) obeying the vortex asymptotic
conditions is straightforward:
\[
\phi _1(y)=C_1(\beta _1)\exp \left[ \frac y{\lambda _{J1}}\right] -C_2(\beta
_1)\exp \left[ \frac y{\lambda _2}\right] ,
\]
\begin{equation}  \label{4.41}
\phi _2(y)=C_1(\beta _1)\exp \left[ \frac y{\lambda _{J1}}\right] +C_2(\beta
_1)\exp \left[ \frac y{\lambda _2}\right] ,\quad y\ll -\lambda _{J1},
\end{equation}
where $C_1(\beta _1)$,$C_2(\beta _1)>0$ are constants with respect to $y$,
parameterized by the initial value $\left| \beta _1\right| $. The
asymptotics for $y\gg \lambda _{J1}$ can be obtained from (\ref{4.41}) by
the use of the symmetry relations (\ref{1.68}):
\[
\phi _1(y)=-C_1(\beta _1)\exp \left[ -\frac y{\lambda _{J1}}\right]
+C_2(\beta _1)\exp \left[ -\frac y{\lambda _2}\right] ,
\]
\begin{equation}  \label{4.42}
\phi _2(y)=2\pi -C_1(\beta _1)\exp \left[ -\frac y{\lambda _{J1}}\right]
-C_2(\beta _1)\exp \left[ -\frac y{\lambda _2}\right] ,\quad y\gg \lambda
_{J1},
\end{equation}
Expressions (\ref{4.41}), (\ref{4.42}) illustrate a very important property
of single-vortex solutions in multilayer structures: Their spatial
dependence is characterized, in general, by $N-1$ different length scales,
which agrees with the conclusions of Ref. \onlinecite{KW97}.

\subsection{A 3-junction stack ($N=4$)}

The consideration of all solutions to (\ref{1.34}) in the case $N\geq 4$,
including thermodynamically unstable incoherent vortex configurations,
requires the use of $\left( N-1\right) \geq 3$ second-order nonlinear
differential equations, or, equivalently, of one nonlinear equation of order
$2\left( N-1\right) \geq 6$. Therefore, for the sake of mathematical
simplicity, from now on we concentrate only on Meissner solutions and
topological vortex-plane solutions, both obeying the symmetry (\ref{1.51}).
(The analysis of incoherent vortex configurations for $N\geq 4$ can be done
using the algorithm worked out in the previous subsection.)

For $N=4$, the Meissner solutions and the vortex-plane solutions satisfy the
relation $\phi _3(y)=\phi _1(y)$. The relevant two equations for $\phi _1(y)$
and $\phi _2(y)$ can be given the form
\[
\frac{d^4\phi _2}{dy^4}+\frac{2+\epsilon ^2}{\epsilon ^2}\sin \phi _2\left(
\frac{d\phi _2}{dy}\right) ^2-\frac{2+\epsilon ^2}{\epsilon ^2}\cos \phi _2
\frac{d^2\phi _2}{dy^2}
\]
\[
-\frac{\frac 14\left[ \frac{2+\epsilon ^2}{\epsilon ^2}\sin \phi _2-\frac{%
d^2\phi _2}{dy^2}\right] \left[ \left( 2+\epsilon ^2\right) \cos \phi _2
\frac{d\phi _2}{dy}-\epsilon ^2\frac{d^3\phi _2}{dy^3}\right] ^2}{1-\frac 14%
\left[ \left( 2+\epsilon ^2\right) \sin \phi _2-\epsilon ^2\frac{d^2\phi _2
}{dy^2}\right] ^2}
\]
\begin{equation}  \label{4.43}
+\sqrt{1-\frac 14\left[ \left( 2+\epsilon ^2\right) \sin \phi _2-\epsilon ^2%
\frac{d^2\phi _2}{dy^2}\right] ^2}\left[ \frac{\left( 2+\epsilon ^2\right)
^2-2}{\epsilon ^4}\sin \phi _2-\frac{2+\epsilon ^2}{\epsilon ^2}\frac{%
d^2\phi _2}{dy^2}\right] =0,
\end{equation}
\[
\phi _1=\arcsin \frac 12\left[ \left( 2+\epsilon ^2\right) \sin \phi
_2-\epsilon ^2\frac{d^2\phi _2}{dy^2}\right] ,\quad -\frac \pi 2\leq \phi
_1\leq \frac \pi 2,
\]
\[
\phi _1=-\arcsin \frac 12\left[ \left( 2+\epsilon ^2\right) \sin \phi
_2-\epsilon ^2\frac{d^2\phi _2}{dy^2}\right] +\pi ,\quad \frac \pi 2<\phi
_1<\pi ,
\]
\begin{equation}  \label{4.44}
\phi _1=-\arcsin \frac 12\left[ \left( 2+\epsilon ^2\right) \sin \phi
_2-\epsilon ^2\frac{d^2\phi _2}{dy^2}\right] -\pi ,\quad -\pi <\phi _1<-%
\frac \pi 2.
\end{equation}
Note that Eqs. (\ref{4.43}), (\ref{4.44}) have the same mathematical
structure as equations for single vortices in the case of a double-junction
stack (\ref{4.34}), (\ref{4.35}). According to (\ref{1.54}), the
superheating (penetration) field for $N=4$ is
\[
H_s=\frac{\sqrt{3\left[ \left( 2+\epsilon ^2\right) ^2-2\right] }}{\epsilon
\sqrt{10+3\epsilon ^2}}.
\]
The local field in the homogeneous Meissner state, according to (\ref{1.33}%
), is given by the relations
\[
H_1=H_3=\frac{\left( 2+\epsilon ^2\right) H}{\left( 2+\epsilon ^2\right) ^2-2%
},\quad H_2=\frac{2H}{\left( 2+\epsilon ^2\right) ^2-2}.
\]

\subsubsection{The Meissner solution for $0<H\ll H_s$}

In fields $0<H\ll H_s$, the general criterion of the existence of Meissner
solutions (\ref{1.53}) takes the form
\begin{equation}  \label{4.47}
\frac 16\left[ \alpha _1^2+\frac{\alpha _2^2}2\right] =\frac{H^2 }{H_s^2},
\end{equation}
where $\alpha _1^2\equiv \phi _1^2(-L)\ll 1$, $\alpha _2^2\equiv \phi
_2^2(-L)\ll 1$. Thus, equations (\ref{4.43}), (\ref{4.44}) can be
linearized:
\begin{equation}  \label{4.48}
\frac{d^4\phi _2}{dy^4}-\frac{2\left( 2+\epsilon ^2\right) }{\epsilon ^2}%
\frac{d^2\phi _2}{dy^2}+\frac{\left( 2+\epsilon ^2\right) ^2-2}{\epsilon ^4}%
\phi _2=0,
\end{equation}
\begin{equation}  \label{4.49}
\phi _1=\frac 12\left[ \left( 2+\epsilon ^2\right) \phi _2-\epsilon ^2\frac{%
d^2\phi _2}{dy^2}\right] .
\end{equation}
The Meissner solution of (\ref{4.48}), (\ref{4.49}) in the region $-L\leq
y<-L+R$, obeying the boundary conditions
\[
\frac{d\phi _1}{dy}\left( -L\right) =\frac{d\phi _2}{dy}\left( -L\right)
=2H,
\]
is
\[
\phi _1(y)=-\frac{\epsilon H}{\sqrt{2}}\left[ \frac{\left( \sqrt{2}+1\right)
}{\sqrt{2-\sqrt{2}+\epsilon ^2}}\exp \left[ -\frac{\left( y+L\right) }{%
\lambda _{J1}}\right] +\frac{\left( \sqrt{2}-1\right) }{\sqrt{2+\sqrt{2}%
+\epsilon ^2}}\exp \left[ -\frac{\left( y+L\right) }{\lambda _{J2}}\right] %
\right] ,
\]
\begin{equation}  \label{4.51}
\phi _2(y)=-\epsilon H\left[ \frac{\left( \sqrt{2}+1\right) }{\sqrt{2-\sqrt{2%
}+\epsilon ^2}}\exp \left[ -\frac{\left( y+L\right) }{\lambda _{J1}}\right] -%
\frac{\left( \sqrt{2}-1\right) }{\sqrt{2+\sqrt{2}+\epsilon ^2}}\exp \left[ -%
\frac{\left( y+L\right) }{\lambda _{J2}}\right] \right] ,
\end{equation}
where
\begin{equation}  \label{4.52}
\lambda _{J1}=\frac \epsilon {\sqrt{2-\sqrt{2}+\epsilon ^2}},\quad \lambda
_{J2}=\frac \epsilon {\sqrt{2+\sqrt{2}+\epsilon ^2}}.
\end{equation}
We observe that spatial dependence of the solution (\ref{4.51}) is
characterized by two different Josephson lengths, $\lambda _{J1}$ and $%
\lambda _{J2}$, in agreement with the general consideration of section III.
Moreover, $\phi _1(y),\phi _2(y)<0$, and $\left| \phi _1(y)\right| <\left|
\phi _2(y)\right| $. The values $\alpha _1\equiv \phi _1(-L)$ and $\alpha
_2\equiv \phi _2(-L)$ meet the existence criterion (\ref{4.47}), as they
should. Using (\ref{4.51}), we can derive explicit expressions for all
physical quantities of interest from general formulas of section II. The
Meissner solution in the region $L-R<y\leq L$ can be obtained from (\ref
{4.51}) by means of the substitution $\phi _{1,2}(y)\rightarrow -\phi
_{1,2}(-y)$. In the region $-R<y<R$, we have $\phi _{1,2}(y)\equiv 0$, and $%
H_n(y)=H_n$, as usual.

\subsubsection{The vortex-plane solution}

The vortex-plane solution to (\ref{4.43}), (\ref{4.44}) in the region $%
-R<y<R $ obeys the initial conditions $\alpha _1\equiv \phi _1(0)=\pi $, $%
\alpha _2\equiv \phi _2(0)=\pi $ and $\beta _1\equiv \frac{d\phi _2(0)}{dy}%
=2H_s$, $\beta _2\equiv \frac{d\phi _2(0)}{dy}=2H_s$ [Eq. (\ref{1.57.1})].
By the symmetry (\ref{1.55}), we also have $\frac{d^2\phi _2(0)}{dy^2}=0$.
The initial condition on $\frac{d^3\phi _2}{dy^3}$ is derived from the
relation
\[
\frac{d^3\phi _2(y)}{dy^3}=\frac 1{\epsilon ^2}\left[ \left( 2+\epsilon
^2\right) \cos \phi _2(y)\frac{d\phi _2(y)}{dy}-2\cos \phi _1(y)\frac{d\phi
_1(y)}{dy}\right]
\]
that follows from (\ref{1.34}):
\[
\frac{d^3\phi _2(0)}{dy^3}=-2H_s.
\]
In this way, we arrive at a complete formulation of the initial value
problem for the vortex-plane configuration. In contrast to the single-vortex
problem considered in the previous subsection, the above-derived initial
conditions do not contain any arbitrariness, in full agreement with our
general conclusion in section III about the uniqueness of the vortex-plane
solutions.

Although the vortex-plane problem admits only numerical integration, the
asymptotics in the regions $\left| y\right| \gg \lambda _{J1}$ and $\left|
y\right| \ll \lambda _{J2}$ can be readily obtained. Thus, on the basis of
linearized Eqs. (\ref{4.48}), (\ref{4.49}), we have:
\[
\phi _1(y)=\frac 1{\sqrt{2}}\left[ C_1\exp \left[ \frac y{\lambda _{J1}}%
\right] -C_2\exp \left[ \frac y{\lambda _{J2}}\right] \right] ,
\]
\[
\phi _2(y)=C_1\exp \left[ \frac y{\lambda _{J1}}\right] +C_2\exp \left[
\frac y{\lambda _{J2}}\right] ,\quad y\ll -\lambda _{J1};
\]
\[
\phi _1(y)=2\pi -\frac 1{\sqrt{2}}\left[ C_1\exp \left[ -\frac y{\lambda
_{J1}}\right] -C_2\exp \left[ -\frac y{\lambda _{J2}}\right] \right] ,
\]
\[
\phi _2(y)=2\pi -C_1\exp \left[ -\frac y{\lambda _{J1}}\right] -C_2\exp %
\left[ -\frac y{\lambda _{J2}}\right] ,\quad y\gg \lambda _{J1},
\]
where $C_1$ and $C_2$ are positive constants. The solution in the region $%
\left| y\right| \ll \lambda _{J2}$ is represented by Taylor series
expansions:
\[
\phi _1(y)=\pi +2H_sy-\frac{\left( 1+\epsilon ^2\right) H_s}{3\epsilon ^2}%
y^3+\ldots ,
\]
\[
\phi _2(y)=\pi +2H_sy-\frac 13H_sy^3+\ldots
\]
Finally, the total flux carried by the vortex plane, by (\ref{1.60}), is
\[
\Phi =\pi \frac{\epsilon ^2\left( 10+3\epsilon ^2\right) }{\left( 2+\epsilon
^2\right) ^2-2},
\]
and the exact value of the induced field at $y=0$, by (\ref{1.61}), is
\[
h_1(0)=h_3(0)=\frac{\epsilon ^2\left( 3+\epsilon ^2\right) H_s}{\left(
2+\epsilon ^2\right) ^2-2},\quad h_2(0)=\frac{\epsilon ^2\left( 4+\epsilon
^2\right) H_s}{\left( 2+\epsilon ^2\right) ^2-2}.
\]
(See Fig. 2.)

\subsection{A 4-junction stack ($N=5$)}

For $N=5$, the Meissner solutions and the vortex-plane solutions satisfy the
relation $\phi _4(y)=\phi _1(y)$ and $\phi _3(y)=\phi _2(y)$. The equations
for $\phi _1(y)$ and $\phi _2(y)$ reduce to the form
\[
\frac{d^4\phi _2}{dy^4}+\frac{1+\epsilon ^2}{\epsilon ^2}\sin \phi _2\left(
\frac{d\phi _2}{dy}\right) ^2-\frac{1+\epsilon ^2}{\epsilon ^2}\cos \phi _2
\frac{d^2\phi _2}{dy^2}
\]
\[
-\frac{\left[ \frac{1+\epsilon ^2}{\epsilon ^2}\sin \phi _2-\frac{d^2\phi _2
}{dy^2}\right] \left[ \left( 1+\epsilon ^2\right) \cos \phi _2\frac{d\phi _2
}{dy}-\epsilon ^2\frac{d^3\phi _2}{dy^3}\right] ^2}{1-\left[ \left(
1+\epsilon ^2\right) \sin \phi _2-\epsilon ^2\frac{d^2\phi _2}{dy^2}\right]
^2}
\]
\begin{equation}  \label{4.56}
+\sqrt{1-\left[ \left( 1+\epsilon ^2\right) \sin \phi _2-\epsilon ^2\frac{%
d^2\phi _2}{dy^2}\right] ^2}\left[ \frac{\left( 2+\epsilon ^2\right) \left(
1+\epsilon ^2\right) -1}{\epsilon ^4}\sin \phi _2-\frac{2+\epsilon ^2}{%
\epsilon ^2}\frac{d^2\phi _2}{dy^2}\right] =0,
\end{equation}
\[
\phi _1=\arcsin \left[ \left( 1+\epsilon ^2\right) \sin \phi _2-\epsilon ^2
\frac{d^2\phi _2}{dy^2}\right] ,\quad -\frac \pi 2\leq \phi _1\leq \frac \pi %
2,
\]
\[
\phi _1=-\arcsin \left[ \left( 1+\epsilon ^2\right) \sin \phi _2-\epsilon ^2
\frac{d^2\phi _2}{dy^2}\right] +\pi ,\quad \frac \pi 2<\phi _1<\pi ,
\]
\begin{equation}  \label{4.57}
\phi _1=-\arcsin \left[ \left( 1+\epsilon ^2\right) \sin \phi _2-\epsilon ^2%
\frac{d^2\phi _2}{dy^2}\right] -\pi ,\quad -\pi <\phi _1<-\frac \pi 2.
\end{equation}
Mathematical structure of Eqs. (\ref{4.56}), (\ref{4.57}) is analogous to
that of Eqs. (\ref{4.43}), (\ref{4.44}) of the 3-junction stack, which
allows us to skip some details in what follows. According to (\ref{1.54}),
the superheating (penetration) field for $N=5$ is
\[
H_s=\frac{\sqrt{2\left[ \left( 2+\epsilon ^2\right) \left( 1+\epsilon
^2\right) -1\right] }}{\epsilon \sqrt{5+2\epsilon ^2}}.
\]
The local field in the homogeneous Meissner state, according to (\ref{1.33}%
), is
\[
H_1=H_4=\frac{\left( 1+\epsilon ^2\right) H}{\left( 2+\epsilon ^2\right)
\left( 1+\epsilon ^2\right) -1},\quad H_2=H_3=\frac H{\left( 2+\epsilon
^2\right) \left( 1+\epsilon ^2\right) -1}.
\]

\subsubsection{The Meissner solution for $0<H\ll H_s$}

For the fields $0<H\ll H_s$, the Meissner solution in the region $-L\leq
y<-L+R$ has the form
\[
\phi _1(y)=-\frac{\epsilon H}{2\sqrt{5}}\left[ \frac{\left( \sqrt{5}%
-1\right) \left( 3+\sqrt{5}\right) }{\sqrt{\frac{3-\sqrt{5}}2+\epsilon ^2}}%
\exp \left[ -\frac{\left( y+L\right) }{\lambda _{J1}}\right] +\frac{\left(
\sqrt{5}+1\right) \left( 3-\sqrt{5}\right) }{\sqrt{\frac{3+\sqrt{5}}2%
+\epsilon ^2}}\exp \left[ -\frac{\left( y+L\right) }{\lambda _{J2}}\right] %
\right] ,
\]
\[
\phi _2(y)=-\frac{\epsilon H}{\sqrt{5}}\left[ \frac{\left( 3+\sqrt{5}\right)
}{\sqrt{\frac{3-\sqrt{5}}2+\epsilon ^2}}\exp \left[ -\frac{\left( y+L\right)
}{\lambda _{J1}}\right] -\frac{\left( 3-\sqrt{5}\right) }{\sqrt{\frac{3+
\sqrt{5}}2+\epsilon ^2}}\exp \left[ -\frac{\left( y+L\right) }{\lambda _{J2}}%
\right] \right] ,
\]
where
\begin{equation}  \label{4.61}
\lambda _{J1}=\frac \epsilon {\sqrt{\frac{3-\sqrt{5}}2+\epsilon ^2}},\quad
\lambda _{J2}=\frac \epsilon {\sqrt{\frac{3+\sqrt{5}}2+\epsilon ^2}}.
\end{equation}
By comparison with $\lambda _{J1}$, $\lambda _{J2}$ of the 3-junction case,
Eq. (\ref{4.52}), the Josephson lengths given by (\ref{4.61}) are larger,
which illustrates a general tendency: Josephson lengths increase with
increasing $N$. As in the 3-junction case, $\phi _1(y),\phi _2(y)<0$, and $%
\left| \phi _1(y)\right| <\left| \phi _2(y)\right| $. The values $\alpha
_1\equiv \phi _1(-L)$ and $\alpha _2\equiv \phi _2(-L)$ satisfy the
existence criterion (\ref{1.53}), as expected.

\subsubsection{The vortex-plane solution}

The vortex-plane solution to (\ref{4.56}), (\ref{4.57}) in the region $%
-R<y<R $ is characterized by the initial conditions $\alpha _1\equiv \phi
_1(0)=\pi $, $\alpha _2\equiv \phi _2(0)=\pi $ and $\beta _1\equiv \frac{%
d\phi _2(0)}{dy}=2H_s$, $\beta _2\equiv \frac{d\phi _2(0)}{dy}=2H_s$ [Eq. (%
\ref{1.57.1})]. In addition, we have $\frac{d^2\phi _2(0)}{dy^2}=0$ and $%
\frac{d^3\phi _2(0)}{dy^3}=-2H_s$, as in the 3-junction case.

The asymptotics of the vortex-plane solution in the region $\left| y\right|
\gg \lambda _{J1}$ are
\[
\phi _1(y)=\frac 12\left[ \left( \sqrt{5}-1\right) C_1\exp \left[ \frac y{%
\lambda _{J1}}\right] -\left( \sqrt{5}+1\right) C_2\exp \left[ \frac y{%
\lambda _{J2}}\right] \right] ,
\]
\[
\phi _2(y)=C_1\exp \left[ \frac y{\lambda _{J1}}\right] +C_2\exp \left[
\frac y{\lambda _{J2}}\right] ,\quad y\ll -\lambda _{J1};
\]
\[
\phi _1(y)=2\pi -\frac 12\left[ \left( \sqrt{5}-1\right) C_1\exp \left[ -%
\frac y{\lambda _{J1}}\right] -\left( \sqrt{5}+1\right) C_2\exp \left[ -%
\frac y{\lambda _{J2}}\right] \right] ,
\]
\[
\phi _2(y)=2\pi -C_1\exp \left[ -\frac y{\lambda _{J1}}\right] -C_2\exp %
\left[ -\frac y{\lambda _{J2}}\right] ,\quad y\gg \lambda _{J1},
\]
where $C_1,C_2>0$. In the region $\left| y\right| \ll \lambda _{J2}$, we
have:
\[
\phi _1(y)=\pi +2H_sy-\frac{\left( 1+\epsilon ^2\right) H_s}{3\epsilon ^2}%
y^3+\ldots ,
\]
\[
\phi _2(y)=\pi +2H_sy-\frac 13H_sy^3+\ldots
\]
The total flux carried by the vortex plane is
\[
\Phi =2\pi \frac{\epsilon ^2\left( 5+2\epsilon ^2\right) }{\left( 2+\epsilon
^2\right) \left( 1+\epsilon ^2\right) -1},
\]
and the exact value of the induced field at $y=0$ is
\[
h_1(0)=h_4(0)=\frac{\epsilon ^2\left( 2+\epsilon ^2\right) H_s}{\left(
2+\epsilon ^2\right) \left( 1+\epsilon ^2\right) -1},\quad h_2(0)=h_3(0)=
\frac{\epsilon ^2\left( 3+\epsilon ^2\right) H_s}{\left( 2+\epsilon
^2\right) \left( 1+\epsilon ^2\right) -1}.
\]
(See Fig. 2.)

\subsection{The layered-superconductor limit ($N-1\gg 2\left[ \frac 1\protect%
\epsilon  \right] $)}

The limit $N-1\gg 2\left[ \frac 1\epsilon \right] $ ($\left[ \frac 1\epsilon %
\right] $ stands for the integer part of $\frac 1\epsilon $) corresponds to
the situation when a periodic thin-layer S/I structure can be regarded as a
''layered superconductor'' rather than merely a ($N-1$)-Josephson-junction
stack. Indeed, in this limit the superheating (penetration) field, Eq. (\ref
{1.54}), becomes
\[
H_s=1+\frac{\sqrt{1+\frac{\epsilon ^2}4}-\frac \epsilon 2}{\epsilon \left(
N-1\right) }
\]
and for $\left( N-1\right) \rightarrow \infty $ tends to the limiting value
of an infinite layered superconductor $H_{s\infty }=1$.\cite{K99,K00} (We
remind that the lower critical field of an infinite layered superconductor,
according to Refs. \onlinecite{K99,K00} , is $H_{c1\infty }=\frac 2\pi $.) The
total flux carried by a vortex plane, by (\ref{1.60}), in the considered
limit is
\[
\Phi =\pi (N-1)\left[ 1-\frac{2\sqrt{1+\frac{\epsilon ^2}4}-\epsilon }{%
\epsilon \left( N-1\right) }\right] ,
\]
which for $\left( N-1\right) \rightarrow \infty $ tends to the limiting
value of one flux quantum $\Phi _0=\pi $ per vortex, as expected for an
infinite layered superconductor.\cite{K99,K00} Moreover, according to (\ref
{1.33}), for I-layers whose index $n$ satisfies the condition $\left[ \frac 1%
\epsilon \right] \ll n\ll N-1-\left[ \frac 1\epsilon \right] ,$ we have
\[
\left| H_n\right| =\left| H\right| \left( \mu ^n+\mu ^{N-n}\right) \ll
\left| H\right| .
\]
In other words, for $\left| H\right| \leq H_{s\infty }=1$, the inside
junctions exhibit the complete Meissner effect in the region $-R<y<R$ .
Thus, the integer $\left[ \frac 1\epsilon \right] $ determines the number of
junctions near the boundaries $x=0$ and $x=N-1$ that ''feel'' the influence
of the superconductor/vacuum interfaces. Far from the boundaries, i.e. in
the region $\left[ \frac 1\epsilon \right] \ll x\ll N-1-\left[ \frac 1%
\epsilon \right] $, one can apply all the results of the theory of infinite
layered superconductors.\cite{K99,K00}

\section{Discussion}

Within the framework of standard methods of the theory of ordinary
differential equations, we have obtained a complete mathematical description
of Josephson vortices and of the Meissner effect in periodic thin-layer S/I
structures. A rigorous examination of the properties of Eqs. (\ref{1.34})
allowed us to establish the general existence criterion (\ref{1.43}), which
formed a solid basis for our subsequent physical and mathematical
conclusions. One of the most striking physical consequences is that the
derivation of the exact expression for the vortex-penetration field $H_s$,
Eq. (\ref{1.54}), did not require any explicit solution of (\ref{1.34}).
Although explicit analytical solutions to Eqs. (\ref{1.34}) proved to be
possible only in a limited number of cases discussed in section IV,
numerical integration of these equations should pose no problem owing to the
algorithm worked out in the paper.

All the three types of localized solutions obtained in the paper possess a
number of interesting physical and mathematical properties. For example, the
Meissner solutions are characterized by several different Josephson lengths $%
\lambda _{Ji}$ ($\frac N2$ lengths for even $N$, and $\frac{N-1}2$ lengths
for odd $N$). Unfortunately, this important fact was not noticed in previous
publications.\cite{BF92} We think that our result may prove to be useful in
view of the current experimental efforts\cite{MKHLX98,PCXWC00} to verify the
interlayer tunneling model of high-$T_c$ superconductivity\cite
{WHA88,A92,CSAS93,A95,A98} by measuring the $c$-axis penetration depth. [The
penetration of the parallel magnetic field with a distribution of length
scales has been recently observed\cite{KMSW99} in the organic layered
superconductor $\kappa $-(BEDT-TTF)$_2$Cu(NCS)$_2$.]

Mathematically, Josephson vortices, represented by both the vortex-plane
solutions and incoherent vortex solutions, are static sine-Gordon-type
solitons. They satisfy the standard\cite{R82} asymptotic boundary
conditions, Eqs. (\ref{1.56}), (\ref{1.57}), and Eqs. (\ref{1.69})-(\ref
{1.71}). The expression for their self-energy (\ref{1.63}) is a direct
generalization of the well-known expression for a single junction.\cite
{K70,BP82} The thermodynamically stable vortex-plane solutions demonstrate a
profound difference between Josephson-vortex formation in weakly-coupled
multilayer structures and Abrikosov-vortex formation in continuum type-II
superconductors (isotropic or not). The proof of the existence of such
solutions in the general case of ($N-1$)-junction stacks establishes
relationship to the well-know ''coherent mode'' (alias the ''in-phase''
mode) in a double-junction stack\cite{SBP93,SAUK96,GGU96,KW97} and the
recently obtained\cite{K99,K00} vortex-plane solutions in infinite ($%
N=\infty $) layered superconductors. However, in contrast to the latter two
cases, the vortex-plane solutions for $4\leq N<\infty $ are characterized by
several $\lambda _{Ji}$, as the Meissner solutions. Moreover, in contrast to
the case $N=\infty $, the intralayer currents, $J_n$, in the presence of a
vortex plane are not equal to zero, with the exception of $J_{\frac{N-1}2}$
in the case of odd $N$.

The single-vortex solutions are not uniquely determined by the asymptotic
boundary conditions (\ref{1.69})-(\ref{1.71}), as follows from the existence
criterion (\ref{1.75}). Their spatial dependence is characterized, in
general, by $N-1$ length scales. In contrast to the vortex planes, isolated
Josephson vortices cannot penetrate the periodic S/I structure at any $%
\left| H\right| \not =0$ and do not satisfy the original integrodifferential
equations (\ref{1.18}), (\ref{1.30}) minimizing the Gibbs free-energy
functional (\ref{1.1}). Although the self-energy of incoherent vortex
solutions may be lower than the self-energy of a vortex plane, their Gibbs
free energy is positive with respect to the Gibbs free energy of the
Meissner state for $\left| H\right| \leq H_{c1}$ ($H_{c1}$ is the lower
critical field for the vortex-plane solution), as is illustrated by our
analysis of a double-junction stack in section IV. However, incoherent
vortex solutions can be realized in the dynamic regime.\cite{SBP93,KW97} In
layered superconductors, isolated vortices may also emerge as metastable
topological entities owing to pinning by extended defects. We emphasize that
our general conclusions about the form of single-vortex configurations stand
in full agreement with numerical results of Ref. \onlinecite{SBP93}.

Finally, our exact results clearly show that the phase-difference equations (%
\ref{1.34}) do not admit any solutions in the form of Abrikosov-type
vortices, suggested in Refs. \onlinecite{B73,CC90}. As we have already pointed out,
\cite{K99,K00} the hypothesis of Abrikosov-type solutions in infinite
layered superconductors leads to incorrect estimates of the value of the
lower critical field $H_{c1\infty }$. Unfortunately, it seems that some
other theoretical predictions, based on the assumption of the existence of
Abrikosov-type solutions, should also be revised.

\appendix

\section{The solution of the finite difference equation for $H_n(y)$}

Equations (\ref{1.30}) can be regarded as a nonhomogeneous finite difference
equation for $H_n(y)$ with respect to the layer index $n$, subject to
boundary conditions (\ref{1.19}). According to general theory of such
equations,\cite{G59} its solution for $\epsilon <1$ can be represented in
the form
\begin{equation}  \label{b.1}
H_n(y)=h_n(y)+H_n,
\end{equation}
where
\begin{equation}  \label{b.2}
h_n(y)=\frac{\epsilon ^2}2\sum_{m=1}^{N-1}G(n,m)\frac{d\phi _m(y) }{dy}
\end{equation}
is the particular solution of (\ref{1.30}) satisfying the boundary
conditions
\begin{equation}  \label{b.3}
h_0(y)=h_N(y)=0,
\end{equation}
and
\begin{equation}  \label{b.4}
H_n=\frac{H\left( \mu ^{-n}+\mu ^{-N+n}-\mu ^n-\mu ^{N-n}\right) }{\mu
^{-N}-\mu ^N},
\end{equation}
\begin{equation}  \label{b.5}
\mu =1+\frac{\epsilon ^2}2-\epsilon \sqrt{1+\frac{\epsilon ^2}4}<1
\end{equation}
is the solution of the homogeneous form of (\ref{1.30}) (with the zero
right-hand side) meeting the boundary conditions
\begin{equation}  \label{b.6}
H_0=H_N=H.
\end{equation}

The quantities $G(n,m)$ in (\ref{b.2}) are matrix elements of $\left(
N+1\right) \times \left( N+1\right) $ matrix Green's function ${\bf G(}0\leq
n,m\leq N)$. They obey the nonhomogeneous finite difference equation
\begin{equation}  \label{b.7}
G(n+1,m)-\left( 2+\epsilon ^2\right) G(n,m)+G(n-1,m)=-\delta _{n,m}
\end{equation}
($\delta _{n,m}$ is the Kronecker index) with the boundary conditions
\begin{equation}  \label{b.8}
G(0,m)=G(N,m)=0.
\end{equation}
The explicit form of $G(n,m)$ is
\begin{equation}  \label{b.9}
G(n,m)=\frac 1{2\epsilon \sqrt{1+\frac{\epsilon ^2}4}}\left[ \mu ^{\left|
n-m\right| }-\frac{\mu ^n\left( \mu ^{m-N}-\mu ^{N-m}\right) +\mu
^{N-n}\left( \mu ^{-m}-\mu ^m\right) }{\mu ^{-N}-\mu ^N}\right] .
\end{equation}
The following properties of $G(n,m)$ can be easily verified using (\ref{b.7}%
) and (\ref{b.9}):
\begin{equation}  \label{b.11}
G(n,m)=G(m,n),
\end{equation}
\begin{equation}  \label{b.12}
G(n,N-m)=G(N-n,m),
\end{equation}
\begin{equation}  \label{b.10}
G(n,m)>0\text{ for any }1\leq n,m\leq N-1,
\end{equation}
\[
\sum_{m=1}^{N-1}G(n,m)=\frac 1{\epsilon ^2}\left[ 1-G(n,1)-G(n,N-1)\right]
\]
\begin{equation}  \label{b.13}
=\frac 1{\epsilon ^2}\left[ 1-\frac{\mu ^{-n}+\mu ^{-N+n}-\mu ^n-\mu ^{N-n}}{%
\mu ^{-N}-\mu ^N}\right] ,\quad 1\leq n\leq N-1,
\end{equation}
\begin{equation}  \label{b.14}
\sum_{n=1}^{N-1}\sum_{m=1}^{N-1}G(n,m)=\frac 1{\epsilon ^2}\left[ N-1-\frac{2%
\sqrt{1+\frac{\epsilon ^2}4}-\epsilon }\epsilon \frac{1-\mu ^{N-1}}{1+\mu ^N}%
\right] ,
\end{equation}
Note that matrix Green's function for an infinite layered superconductor, $%
{\bf G}_\infty (n,m)$, is determined by the matrix elements
\begin{equation}  \label{b.15}
G_\infty (n,m)=\frac{\mu ^{\left| n-m\right| }}{2\epsilon \sqrt{1+\frac{%
\epsilon ^2}4}},\quad -\infty <n,m<+\infty ,
\end{equation}
that satisfy the summation rule
\[
\sum_{m=1}^{N-1}G_\infty (n,m)=\frac 1{\epsilon ^2}.
\]

Consider now a $\left( N-1\right) \times \left( N-1\right) $ matrix ${\bf
\tilde G}(n,m)$, whose matrix elements are given by the right-hand side of (%
\ref{b.9}) with $1\leq n,m\leq N-1$. Of special physical importance are
positive eigenvalues of ${\bf \tilde G}(n,m)$: They determine characteristic
length scales of localized solutions to (\ref{1.34}). Indeed, in the
asymptotic region where $\frac{d\phi _n(y)}{dy}=o(1)$ and $\phi _n(y)=o(1)$,
equations (\ref{1.37}) can be linearized with the result
\[
\sum_{m=1}^{N-1}G(n,m)\frac{d^2\phi _m(y)}{dy^2}=\frac 1{\epsilon ^2}\phi
_n(y),\quad 1\leq n\leq N-1.
\]
The substitution $\phi _n(y)\propto \exp \left[ \pm \frac y\lambda \right] $
yields
\[
\sum_{m=1}^{N-1}G(n,m)\phi _m(y)=\frac{\lambda ^2}{\epsilon ^2}\phi
_n(y),\quad 1\leq n\leq N-1,
\]
which is exactly an eigenvalue equation for ${\bf \tilde G}(n,m)$, with $%
\frac{\lambda ^2}{\epsilon ^2}$ being a positive eigenvalue.

\section{Proof of the {\bf Lemma}}

By introducing new functions
\[
\psi _1(y)=\phi _1(y),\psi _2(y)=\phi _2(y),\ldots ,\psi _{N-1}(y)=\phi
_{N-1}(y),
\]
\begin{equation}  \label{a.1}
\psi _N(y)=\frac{d\phi _1(y)}{dy},\psi _{N+1}(y)=\frac{d\phi _2(y) }{dy}%
,\ldots ,\psi _{2N-2}(y)=\frac{d\phi _{N-1}(y)}{dy},
\end{equation}
we convert (\ref{1.34}) into an equivalent normal system of $2N-2$
first-order equations
\begin{equation}  \label{a.2}
\frac{d\psi _i(y)}{dy}=F_i\left( \psi _1,\psi _2,\ldots ,\psi _{2N-2}\right)
,\quad 1\leq i\leq 2N-2,
\end{equation}
\[
F_i\left( \psi _1,\psi _2,\ldots ,\psi _{2N-2}\right) \equiv \psi
_{i+N-1},\quad 1\leq i\leq N-1,
\]
\[
F_N\left( \psi _1,\psi _2,\ldots ,\psi _{2N-2}\right) \equiv \frac 1{%
\epsilon ^2}\left[ \left( 2+\epsilon ^2\right) \sin \psi _1-\sin \psi _2%
\right] ,
\]
\[
F_i\left( \psi _1,\psi _2,\ldots ,\psi _{2N-2}\right) \equiv \frac 1{%
\epsilon ^2}\left[ \left( 2+\epsilon ^2\right) \sin \psi _i-\sin \psi
_{i-1}-\sin \psi _{i+1}\right] ,\quad N+1\leq i\leq N-3,
\]
\[
F_{2N-2}\left( \psi _1,\psi _2,\ldots ,\psi _{2N-2}\right) \equiv \frac 1{%
\epsilon ^2}\left[ \left( 2+\epsilon ^2\right) \sin \psi _{N-1}-\sin \psi
_{N-2}\right] ,
\]
subject to initial conditions
\[
\psi _i(y_0)=\alpha _i,\quad 1\leq i\leq N-1,
\]
\begin{equation}  \label{a.3}
\psi _i(y_0)=\beta _{i-N+1},\quad N\leq i\leq 2N-2.
\end{equation}

To prove the statement of the {\bf Lemma}, it is sufficient to observe that
all $F_i\left( \psi _1,\psi _2,\ldots ,\psi _{2N-2}\right) $ are continuous
functions of their arguments for $y\in (-\infty ,+\infty )$ and $\psi _k\in
(-\infty ,+\infty )$ ($1\leq k\leq 2N-2$). Moreover, their partial
derivatives with respect to $\psi _k$ satisfy the relation
\begin{equation}  \label{a.4}
\left| \frac{\partial F_i\left( \psi _1,\psi _2,\ldots ,\psi _{2N-2}\right)
}{\partial \psi _k}\right| \leq \frac{4+\epsilon ^2}{\epsilon ^2}
\end{equation}
for $y\in (-\infty ,+\infty )$ and $\psi _k\in (-\infty ,+\infty )$ ($1\leq
i,k\leq 2N-2$). Thus, the Lipschitz conditions with respect to $\psi _k$ are
met for $y\in (-\infty ,+\infty )$ and $\psi _k\in (-\infty ,+\infty )$ ($%
1\leq k\leq 2N-2$), which immediately guarantees\cite{T61} the existence and
uniqueness of a solution to (\ref{a.2}), satisfying arbitrary initial
conditions (\ref{a.3}), in an arbitrary interval $I=\left[ L_1,L_2\right] $
such that $y_0\in I$. Continuous dependence of the solution on initial data
is a result of continuous dependence of $F_i\left( \psi _1,\psi _2,\ldots
,\psi _{2N-2}\right) $ on their arguments and of the condition (\ref{a.4}).
Infinite differentiability of the solution automatically follows from
infinite differentiability of $F_i\left( \psi _1,\psi _2,\ldots ,\psi
_{2N-2}\right) $ with respect to their arguments.

\section{The upper and the lower bounds for the self-energy of single-vortex
solutions in a double-junction stack}

To determine the upper and the lower bounds of the self-energy of
single-vortex solutions in a double-junction stack, we must find the extrema
of the right-hand side of (\ref{1.63}) with $N=3$ under the normalization
condition
\[
\sum_{n=1}^2\int\limits_{-R}^Rdy\left[ \frac{d\phi _n(y)}{dy}\right] ^2=
\text{const}<\infty .
\]
This leads to the variational principle
\[
\frac \delta {\delta \phi _n(y)}\left[ \epsilon
^2\sum_{k=1}^2\sum_{m=1}^2\int\limits_{-R}^RduG(k,m)\frac{d\phi _k(u)}{du}
\frac{d\phi _m(u)}{du}-\lambda ^2\sum_{k=1}^2\int\limits_{-R}^Rdu\left[
\frac{d\phi _k(u)}{du}\right] ^2\right] =0,
\]
where $\lambda ^2$ is a Lagrange multiplier. Performing the variation with
the use of the boundary conditions $\frac{d\phi _n(\pm R)}{dy}\rightarrow 0$%
, we arrive at the eigenvalue problem for the $2\times 2$ matrix ${\bf
\tilde G}(n,m)$, determined by the matrix elements (\ref{4.20}):
\[
\sum_{m=1}^2G(n,m)\frac{d^2\phi _m(y)}{d^2y}=\frac{\lambda ^2}{\epsilon ^2}
\frac{d^2\phi _n(y)}{d^2y}.
\]
The only two eigenvalues of ${\bf \tilde G}(n,m)$ are $\frac{\lambda _{J1}^2
}{\epsilon ^2}$ and $\frac{\lambda _2^2}{\epsilon ^2}$, with $\lambda _{J1}$
and $\lambda _2$ given by (\ref{4.21}). The larger eigenvalue, $\frac{%
\lambda _{J1}^2}{\epsilon ^2}$, corresponds to the vortex-plane solution
with $\phi _1=\phi _2$ and determines the upper bound (\ref{4.28}) for the
self-energy. The smaller eigenvalue, $\frac{\lambda _2^2}{\epsilon ^2}$,
corresponds to the vortex-antivortex solution with $\phi _1=-\phi _2$ and
determines the lower bound (\ref{4.33}) for the self-energy. Thus, for the
self-energy of the single-vortex solutions, $E_{sv}$, we necessarily have
\[
E_{va}<E_{sv}<E_v\text{.}
\]

\begin{center}
\newpage\ {\bf FIGURE CAPTIONS}
\end{center}

\begin{enumerate}
\item  Fig. 1. The geometry of the problem (schematically). Here $N=12$; $%
\lambda _{J\infty }<<R<L/2$, $\lambda _{J\infty }^{-2}=8\pi ej_{0}p$; $H>0$.

\item  Fig. 2. The distribution of the self-induced field of a vortex plane $%
h(x,y)$ at $y=0$, for $\epsilon =0.5$: a) a 3-junction stack, $H_{s}=1.849$;
b) a 4-junction stack, $H_{s}=1.662$.
\end{enumerate}

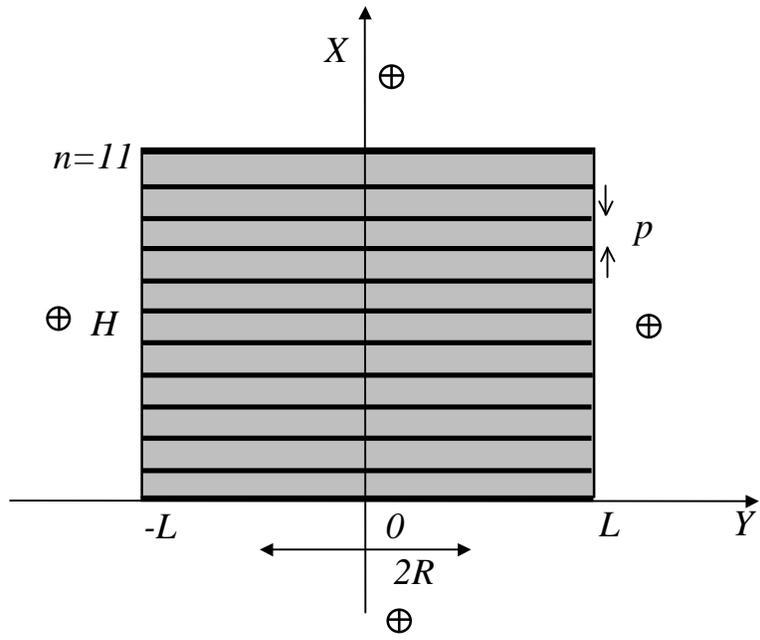
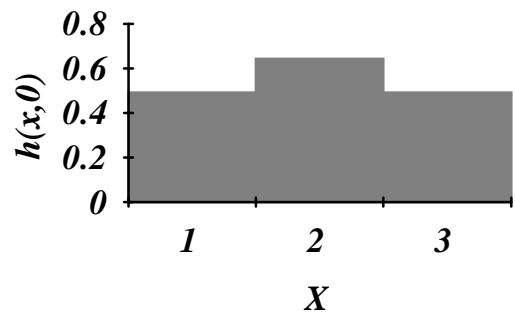
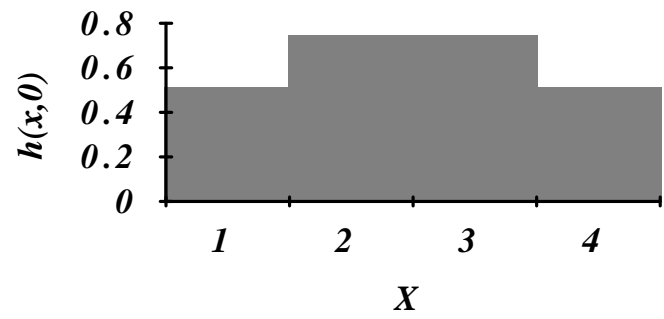
\end{document}